\documentclass[]{nature_mod}

\usepackage{amssymb,amstext,amsmath}
\usepackage{graphicx}
\usepackage{dcolumn}
\usepackage{bm} 
\usepackage{multirow}
\usepackage{bigstrut}
\usepackage{makecell,rotating}
\usepackage{mathrsfs}
\usepackage{booktabs}
\usepackage{threeparttable}
\usepackage{subfigure}
\usepackage{chngpage}
\usepackage{float}
\usepackage{color}
\usepackage{xcolor}

\usepackage{xr}
\usepackage[normalem]{ulem}
\usepackage{color}
	 \definecolor{darkred}{rgb}{0.75,0,0}
	 \definecolor{darkgreen}{rgb}{0,0.5,0}
	 \definecolor{darkblue}{rgb}{0,0,0.75}
  	 \definecolor{darkorange}{rgb}{1,0.9,0.1}
	 \definecolor{dark}{rgb}{0,0,0}

\usepackage{lineno}

\begin{document}
\title{
Optimal network structure for collective performance with strategic information sharing}
\author{Ye Wang$^{1,2}$, Andrea Civilini$^{2,3}$, Anzhi Sheng$^{4,5}$, Xiaojie Chen$^{6,*}$, Long Wang$^{1,7,*}$, \& Vito Latora$^{2,8,9,*}$}
\maketitle
\begin{enumerate}
  \item  Center for Systems and Control, College of Engineering, Peking University, Beijing 100871, China
  \item School of Mathematical Sciences, Queen Mary University of London, London E1 4NS, United Kingdom
  \item Sorbonne Universit\'e, Paris Brain Institute (ICM), CNRS UMR7225, INRIA Paris, INSERM U1127, H\^opital de la Piti\'e Salp\^etri\`ere, AP-HP, Paris 75013, France
  \item Division of Decision and Control Systems, School of Electrical Engineering and Computer Science, KTH Royal Institute of Technology, Stockholm 10044, Sweden
  \item Department of Mathematics, KTH Royal Institute of Technology, Stockholm 10044, Sweden
  \item School of Mathematical Sciences, University of Electronic Science and Technology of China, Chengdu 611731, China
  \item Center for Multi-Agent Research, Institute for Artificial Intelligence, Peking University, Beijing 100871, China
  \item Dipartimento di Fisica ed Astronomia, Universit\`a di Catania and INFN, Catania I-95123, Italy
  \item Complexity Science Hub Vienna, Vienna A-1080, Austria
   \item[] Corresponding authors: xiaojiechen@uestc.edu.cn, longwang@pku.edu.cn, \\ v.latora@qmul.ac.uk
\end{enumerate}


\newpage

\begin{abstract} 
Information sharing between individuals is crucial to improve performance in collective tasks. 
However, in a competitive world, individuals may be reluctant to share information with the others, and it is still unclear how the presence of strategic behaviors affects 
the collective performance of a group. 
In this study, we introduce an evolutionary game modeling the dynamics of individual behaviors in a collective estimation task. The individuals are organized in a network and have to guess the distribution of ball colors in a box. Each of them samples a given number of balls and can strategically decide whether to share or not this information with its neighbors. We develop a framework that allows to investigate analytically how the collective performance depends on the network structure. 
We find that the optimal network results from a trade-off between the sharing rate and the way the information is integrated in the network. 
We further reveal that there exists an intermediate average degree for each type of network maximizing the collective performance. 
In addition to the uniform case, we consider the case of non-homogeneous allocations of the number of individual samples, showing that the largest collective performance is obtained when the number of ball extracted by an individual is inversely proportional to its degree. 

\end{abstract}
\section{Introduction}

Collective intelligence, which can be defined as the ability of a group to perform a broad range of tasks, lies at the heart of disciplines ranging from biology and ecology to social sciences and AI\cite{Woolley2010, Sayin2025,Riedl2021,Peeters2021,Flack2018,Kameda2022,Strandburg-Peshkin2015}. Among the collective tasks, estimation tasks, namely those that require individuals to estimate the true value of a quantity, are widespread and particularly important\cite{Almaatouq2020,Centola2022}. In such tasks, individuals typically have access to limited information, resulting in inaccurate estimates when acting alone. However, individuals are usually organized in a network and can achieve higher estimation accuracy through prosocial interactions\cite{Farrell2011,Gurcay2015}, when they are keen to share information with their neighbors, therefore improving group performance\cite{Farrell2011,Jayles2017}. Notably, the manner in which social interactions influence group performance remains poorly understood from a bottom-up perspective, which is a central concern in the research field of collective intelligence\cite{Kameda2022,Centola2022,Sehwag2025,Barfuss2025}.

Previous studies have proposed various measures to quantify group performance from different perspectives\cite{Jayles2017,Farrell2011,Latora2001,VanDolder2017,Tian2023,Tian2022,Masuda2015,Han2024,Han2026}. Among these, the average performance of individuals is a widely used and particularly important indicator\cite{Jayles2017,Farrell2011}, as it captures overall performance at the individual level and reflects the performance of the entire group. Investigations based on average performance of individuals have shown that social communication and information sharing enable participants to achieve highly accurate estimates, even when their initial judgments are biased or unreliable\cite{Almaatouq2020,Farrell2011,Gurcay2015,Jayles2017}. Intuitively, when all individuals are willing to sharing their own information with their neighbors, the average estimation performance can be significantly improved. This is because such a state of full sharing can maximize the efficiency of information transmission in the population. In this context, previous works have largely focused on the effects of network structures on the information transmission and looked for the most efficient networks in terms of information spreading\cite{Latora2001,Latora2007,Vragovic2005}. It is found that a well-mixed population works as an efficient network, allowing information to spread widely and resulting in high average estimation performance\cite{Latora2001,Centola2022}. 

However, individuals are not always willing to share their information during social interactions. Indeed, when an individual shares information with a neighbor on the network, it benefits the recipient but may incur a cost to itself\cite{Leonard2022,Su2022}. This cooperative behavior may not have an evolutionary advantage over non-sharing behavior\cite{Leonard2022,Su2022,nowak2006evolutionary,Broom2019}. Thus, whether to share or not information with neighbors can be regarded as a decision-making process with strategic choices. In such a scenario, the efficacy of information transmission for promoting group performance largely depends on the sharing rate in a population of individuals with strategic behaviors. Therefore, some recent efforts have concentrated on promoting the cooperation rate in order to favor collective performance\cite{Duong2021,Perc2010,Perc2013,Meng2024,Barfuss2025,Sehwag2025,Szolnoki2013,Tchernichovski2023,Duong2023,Masuda2003}. It has been found that in structured populations, individuals will choose to share information with neighbors by forming cooperative clusters to resist invasion from non-sharing strategic behavior, thus leading to a high cooperation (sharing) level\cite{Ohtsuki2006,Allen2017}.

Although it is meaningful to consider the dynamics of 
information sharing among the individuals involved in a collective estimation task, related studies focusing on sharing behaviors\cite{Leonard2022,Su2022} may not capture the essence of promoting collective estimation performance. It is worth mentioning that many works have been investigating the effects of network structures on promoting sharing behaviors. Such studies have shown that networks with a low average degree tend to favor the emergence of sharing behaviors\cite{Ohtsuki2006,Fotouhi2019,Sheng2023}, but they may not transmit information efficiently\cite{Latora2001} throughout the population, which is detrimental to collective performance. In contrast, networks with a high average degree, such as well-mixed populations, disfavor sharing behavior from an evolutionary perspective\cite{Allen2017,nowak2006evolutionary}, but they can enhance collective performance in the full-sharing state by enabling rapid information transmission\cite{Latora2001,Centola2022} across the population. Thus a discrepancy may exist between average estimation performance and sharing rate, when the dynamics of information sharing
is considered in networked populations. 
Then, a natural yet unexplored question arises accordingly: what network structure does maximize group estimation performance in populations with individual strategic sharing behaviors? To the best of our knowledge, no study to date has addressed this issue.

Here we develop a mathematical framework to study group performance in the case when the individuals are organized  in a network, and they 
can strategically choose whether to share or not their information with the  neighbors. The collective task they need to solve is 
the estimation of the true distributions of ball colors in a box, from a set of partial samplings.
In each round, individuals independently sample a finite number of balls and decide whether to share their observations with neighbors. They then form estimates of the true distribution based on both their individual and shared information, receive payoffs proportional to the estimation accuracy, and update their information-sharing strategies accordingly. Using a game-theoretical approach\cite{Ohtsuki2006,Allen2017,Wang2022,Broom2010,broom2013game}, we provide an analytical framework to model the evolutionary dynamics between sharing (cooperative) and non-sharing (non-cooperative) strategies on networks, revealing a fundamental trade-off: networks that achieve high performance under full sharing often suppress the emergence of sharing behavior, whereas networks that promote sharing may fail to achieve high collective performance under the full-sharing state. Consequently, the optimal network structure results from a balance between these two competing effects, rather than being determined by a single factor. Beyond capturing the evolutionary dynamics of sharing strategies, our analytical approach 
provides a practical 
and efficient way to identify the optimal network from an arbitrary set, without requiring extensive numerical simulations. Applying this approach, we find that the average degree of the network has a nonlinear effect on the average individual error, allowing us to analytically determine the optimal degree in random regular graphs. Our framework can also accommodate more general scenarios, including heterogeneous initial sample allocations, where an allocation inversely proportional to degree can yield superior performance. Overall, this work establishes a unifying framework for understanding how network structure shapes collective performance in natural and social systems, providing new perspectives on enhancing group-level outcomes.

\section{Results}
\subsection{Model of collective estimation} 
We consider a structured population of $N$ individuals participating in a collective estimation task. The population structure is described by a graph $\mathbf{G}=\left[w_{ij}\right]_{i,j=1}^{N}$, whose nodes represent individuals and edges indicate the relationships between them. Here $w_{ij}=1$ if an edge exists between individuals $i$ and $j$, otherwise $w_{ij}=0$. The collective estimation task consists in guessing the true distribution of ball colors in a box containing $n$ distinct colors\cite{Friedman1949}. The true distribution of the $n$ colors is $\mathbf{p}=\left\{p_1, \dots, p_{\alpha}, \dots, p_n\right\}$, where $p_{\alpha}$ is the fraction of balls with the ${\alpha}$-th color in the box and $\sum_{{\alpha}=1}^np_{\alpha} =1$. At each round of the collective estimation task,  each individual $i$ is asked to perform $s_i$ independent draws of balls with replacement from the box, accordingly obtaining $s_i$ initial observations of the colors in the box (Fig.~1A). After performing their independent draws, the individuals are allowed to engage in pairwise interactions with their neighbors in the graph. During the interaction process, each individual can adopt one of two strategies: sharing its observations ($S$ strategy) or withholding them ($W$ strategy). When choosing to share, individual $i$ needs to pay a cost $c_i$ to transmit its information to one neighbor, who will benefit from $i$'s information; otherwise, $i$ can decide to not share its observations, thus not paying any cost, but can still receive information from sharing neighbors (Fig.~1B). Note that such sharing choice is a cooperative behavior, and rational individuals would prefer to withhold their observations according to classical game theory\cite{Leonard2022,Su2022}. Based on both its own observations and on the information obtained through social interactions with all the neighbors, each individual $i$ collects $S_i$ samples, thus making a final estimate of the distribution of balls with $n$ colors in the box, 
$
\mathbf{\hat{p}}^{(i)} = \left\{\hat{p}_1^{\left(i\right)}, \hat{p}_2^{\left(i\right)}, \dots, \hat{p}_n^{\left(i\right)}\right\}
$,
where $\hat{p}_{\alpha}^{\left(i\right)}$ for $\alpha=1,\cdots,n$ denotes $i$'s estimate of the fraction of balls of color
$\alpha$. 

The estimation performance of individual $i$ is assessed by the estimation error, $E_i$, which measures the deviation of the estimate of $i$ from the true distribution and is given by\cite{Centola2022}
\begin{equation}
E_{i}=\|\mathbf{\hat{p}}^{(i)} - \mathbf{p}\|^2=\sum_{\alpha=1}^{n}\left(\hat{p}_{\alpha}^{(i)}-p_{\alpha}\right)^2,    
\end{equation} where a smaller error indicates better performance. 
To mitigate stochastic fluctuations in individual performance, 
we consider multiple independent realizations of each round of the process, 
and quantify $E_i$ as the average estimation error of individual $i$ over the different realizations.
%
Then, each individual $i$ receives a reward 
$f\left(E_{i}\right)$, which is a function inversely proportional to the estimation error $E_{i}$. Accordingly, the total payoff of an individual is obtained by combining this reward with all the costs of sharing with neighbors. Specifically, let $x_i$ be the strategy of individual $i$ in the current round, where $x_i =1$ indicates that $i$ is a sharer ($S$) and $0$ otherwise ($W$). Therefore, the total payoff of $i$, $u_i$, is written as:
\begin{equation}
u_{i}=f\left(E_{i}\right)-w_i c_i x_i, 
\end{equation}
where $w_i=\sum_{j=1}^{N}w_{ji}$ is the degree of node $i$, namely 
the number of neighbors of individual $i$. For the reward function $f$, we employ the classical form $f\left(E_{i}\right)=R/E_{i}$, where $R$ denotes a baseline reward. The total payoff $u_i$ determines the fitness $F_{i}=\exp{\left(\theta u_i\right)}$ of individual $i$, with $\theta$ being the selection strength, a quantity widely used in studies of evolutionary dynamics\cite{Wu2010,nowak2006evolutionary}.

Finally, we assume that individuals can observe the strategic behaviors and fitness of their neighbors, and have the opportunity to update their strategies of sharing or not their 
information at the end of each round. 
Specifically, the strategy update is implemented as a Death-Birth process\cite{nowak2006evolutionary,Ohtsuki2006,Allen2017}. Namely, a randomly selected individual $i$ updates its strategy by imitating the strategy of a neighbor, with a probability proportional to its fitness. The probability that $i$ adopts the strategy of $j$ is given by: 
\begin{equation}
P_{j\rightarrow i} =\frac{1}{N}\frac{F_j w_{ji}}{\sum_{m=1}^{N}F_m w_{mi}}.
\end{equation}
In addition to imitation of strategic behaviors, we assume that rare explorations (errors) occur during the strategy updating process\cite{Fudenberg2006,Wang2022,Hauert2007}. Specifically, with an exploration rate (also called error rate) $\mu$, the selected individual $i$ updates its strategy by adopting another strategy at random, otherwise individual $i$ imitates the strategy of neighbor $j$ according to the procedure mentioned above. Since explorations are rare, we consider the evolutionary model in the limit of low exploration rates. Note that this mechanism of rare random exploration prevents the population from becoming occasionally trapped in a homogeneous state and enables a more accurate theoretical evaluation of collective performance.

\begin{figure*}[!h]
\centering
\includegraphics[width=16cm]{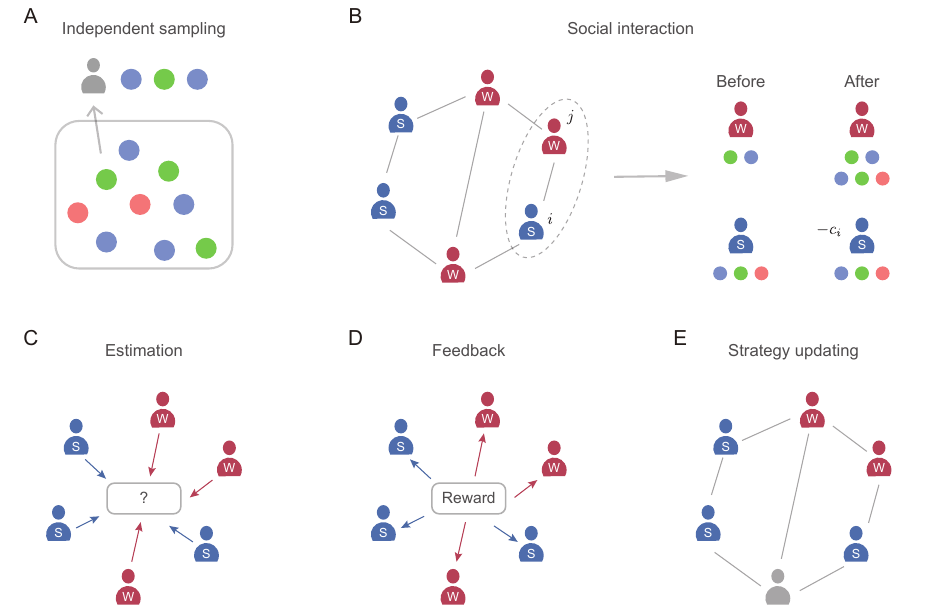}
\caption{\textbf{Illustration of the collective estimation task framework.} (A) At each round, every participant $i$ independently draws $s_i$ samples with replacement from the box and records the colors of the sampled balls. The number of samples drawn may vary across individuals. Hence the specific samples obtained can differ due to individual sampling variability, although the color distribution of the balls in the box is fixed. (B) Social interaction process. During social interactions, each individual may choose between two strategies: sharing ($S$) or withholding ($W$) its information. Consider the interaction between sharer $i$ (blue) and non-sharer $j$ (red). As a sharer, $i$ pays a cost $c_i$ to transmit its sample information to $j$, so that $j$ obtains additional samples. However, since $j$ is a non-sharer, no information is transmitted from from $j$ to $i$. (C) After social interactions, each participant submits an estimate of the color distribution in the box based on both its own samples and additional samples obtained from sharing neighbors. Here, the larger the number of total samples, the smaller the expected estimation error. (D) Individuals receive rewards according to their estimation errors, with smaller errors corresponding to higher rewards. (E) Based on the total payoff (rewards minus costs), a randomly selected individual (gray) updates its strategy by imitating the strategy of one of its neighbors. The population then proceeds to the next round of the estimation task.}
\label{Illustration}
\end{figure*}

Collective performance is typically assessed using two different quantities, the error of the group’s average estimate and the average of the individual estimation error\cite{Jayles2017}. Previous studies have shown that social interactions through information aggregation do not significantly influence the error of the group’s average estimate\cite{Lorenz2011}. Conversely, information aggregation can have a profound effect on the second quantity, the average individual error\cite{Farrell2011}, which serves as a key indicator of collective performance under social interactions. Thus, in this work we focus on the average individual error to quantify collective performance\cite{Farrell2011,Jayles2017}. Accordingly, the average individual error of a graph $\mathbf{G}$, $E_{\mathbf{G}}$, is given by: 
\begin{equation}
 E_{\mathbf{G}} = \frac{1}{N} \sum_{i=1}^{N} E_{i},
\end{equation}
and captures the collective performance across the network. 
Notably, a smaller average individual error indicates a higher accuracy of collective estimation, and corresponds to a higher collective performance. 

\subsection{Theoretical analysis framework}

We focus on the expected collective performance over a long time horizon, so that we can capture the overall dynamics while minimizing short-term fluctuations.
Note that under rare exploration rates, the population evolves toward one of two absorbing homogeneous states, either all sharers or all non-sharers, and the time spent in intermediate states is negligible over the long term\cite{Fudenberg2006,Wang2022,Hauert2007}. This allows us to approximate the evolutionary dynamics of strategic behaviors by means of an embedded Markov chain whose states correspond to the two homogeneous states of the population\cite{Fudenberg2006}. The stationary distribution of the population can be thus used to measure the average
fraction of time that the population spends in each of the homogeneous states. We respectively use $t_S$ ($t_W$) to denote the average fraction of time that the population spends in the full-sharing (non-sharing) state. Thus we have $t_S=1-t_W$ and we emphasize that the quantity $t_S$ characterizes the sharing rate during the whole evolutionary process. Consequently, the expected average individual error is (see Supplementary Section 1): 
\begin{equation}
\mathbb{E}\left[E_{\mathbf{G}}\right]={t_S}\mathbb{E}\left[E_{\mathbf{G}}^{S}\right]+{t_W}\mathbb{E}\left[E_{\mathbf{G}}^{W}\right]={t_S}\left(\mathbb{E}\left[E_{\mathbf{G}}^{S}\right]-\mathbb{E}\left[E_{\mathbf{G}}^{W}\right]\right)+\mathbb{E}\left[E_{\mathbf{G}}^{W}\right].
\label{eqs:expected}
\end{equation}
Here, $\mathbb{E}\left[E_{\mathbf{G}}^{S}\right]$ is the expected average individual error in the full-sharing state, which is given by (see Supplementary Section 1)
\begin{equation}
   \mathbb{E}\left[E_{\mathbf{G}}^{S}\right]= \sum_{i=1}^{N}\frac{1-\sum_{\alpha=1}^{n}p_{\alpha}^{2}}{N \left(s_i + \sum_{l=1}^{N} w_{li} s_l\right)},
   \label{exp_s}
\end{equation}
where $\sum_{l=1}^{N} w_{li} s_l$ denotes the number of samples individual $i$ obtained through social interactions. In addition, $1 - \sum_{\alpha=1}^{n} p_{\alpha}^{2}$ follows the classical definition of Gini impurity\cite{Jost2006,breiman2017classification}, a measure in which larger values indicate greater uncertainty in the distribution and can be regarded as an indication of a higher task complexity.

Similarly, the expected average individual error in the non-sharing state, $\mathbb{E}\left[E_{\mathbf{G}}^{W}\right]$, is given by (see Supplementary Section 1)
\begin{equation}
   \mathbb{E}\left[E_{\mathbf{G}}^{W}\right]= \sum_{i=1}^{N}\frac{1-\sum_{\alpha=1}^{n}p_{\alpha}^{2}}{N s_i}.
   \label{exp_w}
\end{equation}
Indeed, the average individual error in the non-sharing state is independent of the underlying population structure. Therefore, the collective performance under strategy evolution on networks of 
Eq.~(\ref{eqs:expected}) primarily depends on two key quantities: $\mathbb{E}\left[E_{\mathbf{G}}^{S}\right]$ (the expected average error under full sharing) and $t_S$ (the sharing rate). 

Additionally, we can approximate the sharing rate ($t_S$) and non-sharing rate ($t_W$) using the evolutionary game-theoretic approach proposed by Fudenberg and Imhof\cite{Fudenberg2006} as follows (see Supplementary Section 2)
\begin{equation}
\begin{cases}
t_S = \frac{\rho_S}{\rho_S + \rho_W},\\
t_W = \frac{\rho_W}{\rho_S + \rho_W}.
\end{cases} 
\label{sta}
\end{equation}
Here, $\rho_S$ is the fixation probability of a single sharer to take over the population of non-sharers, while $\rho_W$ is the fixation probability of a single non-sharer to take over the population of sharers\cite{Allen2017,McAvoy2021,Broom2008}.

According to Eqs.~(\ref{eqs:expected})-(\ref{sta}), the expected average individual error is then written as
\begin{equation}
\begin{aligned}
\mathbb{E}\left[E_{\mathbf{G}}\right]=\frac{1-\sum_{\alpha=1}^{n}p_{\alpha}^{2}}{N\left(\rho_S+\rho_W\right)}\sum_{i=1}^{N}\left(\frac{\rho_S}{s_i+\sum_{l=1}^{N}w_{li}s_l}+\frac{\rho_W}{s_i}\right)
\end{aligned}.
\label{eqs:cet2}
\end{equation}
For every given network structure, we can then 
characterize the collective performance of the population 
by using Eq.~(\ref{eqs:cet2}) to calculate the corresponding expected average individual error. 
In particular, we can derive analytical approximations for the fixation probabilities $\rho_S$ and $\rho_W$ under weak selection (i.e., $\theta \ll 1$)\cite{Ohtsuki2006,Allen2017,McAvoy2021}, thereby obtaining an explicit expression for the expected average individual error (see Methods and Supplementary Section 2). This also allows us to compare the collective estimation performances in various networks, and then identify the features of population structures that most effectively enhance collective performance. 



\subsection{The network multifaceted effects on collective performance}

To investigate the role of the network in shaping the performance of collective estimation, we consider all the connected, unweighted graphs of five individuals\cite{Allen2017} as a typical example. Specifically, we calculate the average individual error for each of the $21$ different graphs of five nodes, thus identifying the structure that minimizes this error for the highest collective performance. When each individual draws the same number of balls and each sharer pays the same sharing cost (i.e., $s_i=s$ and $c_i=c$ for $i=1,\cdots,N$), we find that the ring graph exhibits the smallest average individual error and is therefore the optimal for collective performance (Fig.~2A). 
Besides, our theoretical predictions are in agreement with numerical simulations. However, when the population is in the full sharing state, the average individual error is minimized on the complete graph. The complete graph thus maximizes the estimation performance under full sharing (Fig.~2B). Besides, when strategy evolution of information sharing is considered, we find that the line graph is the optimal network that most promotes sharing behaviors. In contrast, the complete graph exhibits the lowest sharing rate, defined as the fraction of time the population spends in the full-sharing state ($t_S$), among all $21$ graphs in this scenario (Fig.~2C). 
We can thus conclude that the results in structured populations with strategy evolution are significantly different 
from those obtained by either considering collective performance in the full sharing state\cite{Latora2001,Almaatouq2020}, or measuring collective performance by the sharing rate only\cite{Wang2022,Hauert2007}.

Notably, the ring graph does not produce the minimal average individual error under full sharing or generate the highest sharing rate under strategy evolution, but relatively high levels of both estimation performance under full sharing and sharing rate can be attained in the network. Thus the ring graph is both ranked in the middle range of the network structures according to these two quantities (Fig.~2D). But it makes a fundamental trade-off between the collective performance under full sharing and the sharing rate, leading to the highest collective estimation performance when strategy evolution is taken into account.

\begin{figure*}[!h]
\centering
\includegraphics[width=16cm]{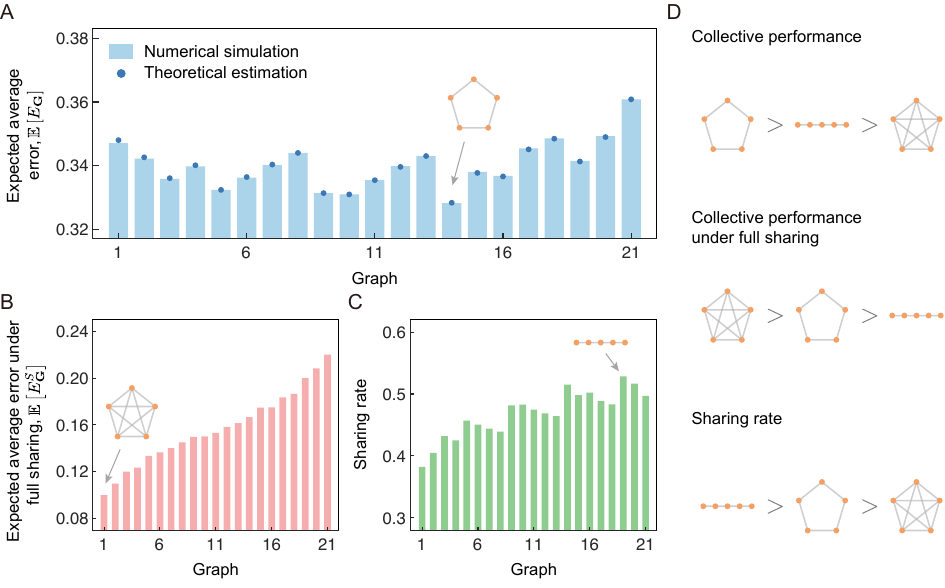}
\caption{\textbf{Collective performance on various networks.} (A) Expected average individual error across all connected, unweighted graphs of size five. Analytical predictions (symbols) of the error are presented to compare with numerical results (bars) from Monte Carlo simulations (see Methods). Among the $21$ graphs, the ring graph achieves the best collective performance, with the lowest average individual error under the given parameter values. (B) Expected average individual error in the full sharing state across the $21$ graphs. When all individuals share information, the complete graph exhibits the lowest average error, attaining the best collective performance. (C) Information sharing rate across the $21$ graphs. The line graph exhibits the highest sharing rate, whereas the complete graph shows the lowest. (D) Comparison among ring, line, and complete graphs of size five. Although the ring graph neither minimizes estimation error under full sharing nor maximizes the sharing rate, it maintains relatively high levels at both, achieving the highest overall collective performance. Here we consider the uniform case in which each individual has identical drawn samples and sharing costs. Parameters: $N=5$, $\theta=0.01$, $R=6$, $n=2$, $p_1=p_2=\frac{1}{2}$, and $s_i=c_i=1$ for $i=1,\cdots,N$.}
\label{Coupled_effects}
\end{figure*}

Furthermore, we find that the optimal network that maximizes collective estimation performance under strategy evolution depends on parameters such as the baseline reward and selection strength (Supplementary Figs.~S1-2). Specifically, different parameter settings affect the performance under full sharing and the sharing rate to different extents, which in turn modulates the strength of the trade-off between these two factors and determines the optimal network. For example, by setting $\theta=0$, our model reduces to the case considered in previous studies that do not account for the social dilemma of information sharing\cite{Centola2022,Almaatouq2020}. In such a case complete graphs are the optimal structures for group performance\cite{Latora2001}. As $\theta$ increases from $0$ to $0.01$, this trade-off becomes more pronounced. In this case, the average individual error increases in complete graphs but decreases in ring and line graphs (Supplementary Fig.~S1). Consequently, the optimal network among the 21 graphs of size 5 shifts from complete graphs, where strategy competition has little effect and information spreading is maximized\cite{Latora2001}, to ring graphs, where the influence of strategy evolution becomes significant (Supplementary Fig.~S2). 


Beyond all graphs with five nodes, we also consider larger networks, such as random regular graphs\cite{newman2010networks, Latora2017} (Supplementary Fig.~S3), demonstrating the effectiveness of our theoretical framework in predicting the average individual error, and confirming 
the significant role of strategy evolution in collective performance. 
Our results reveal that networks with higher average degree, although advantageous for information spreading, do not necessarily enhance collective performance, as they may not promote the evolution of sharing behavior\cite{Ohtsuki2006,Allen2017}. Similar results are observed across other network topologies, including Watts–Strogatz small-world (SW) networks\cite{Watts1998}, Erd\H{o}s--R\'enyi random (ER) networks\cite{erdos1960evolution}, and Barab\'asi--Albert scale-free (SF) networks\cite{Albert2002,Barabasi1999} (Supplementary Fig.~S4), confirming the robustness of our theoretical predictions and analytical results across diverse network structures. Together, our framework provides a comprehensive approach for characterizing collective behavior and the factors that govern collective performance on complex networks.

\subsection{Identifying the optimal network for collective performance}
As discussed above, our analytical framework enables accurate prediction of the average individual error across diverse network topologies. This allows us to identify the optimal network within a given set of networks by computing the expected average error for each topology and selecting the one with the lowest value. Furthermore, it allows to clarify how different network metrics affect 
collective performance. In particular, we expect that the average degree $k$ of a network should play a crucial role in determining the collective performance when the evolution of strategies is taken into account. 
This is because a larger value of $k$ enhances collective performance under full sharing by enabling more efficient information spreading, whereas a smaller $k$ is more favorable for the evolution of sharing behavior\cite{Ohtsuki2006,Allen2017}. 
Notably, our analytical framework enables us to know that the collective estimation performance indeed depends on the average degree and there exists an optimal degree $k^*$ that maximizes collective performance (under weak selection) when all individuals have identical sampling rates and each sharer pays the same sharing cost. An analytical expression for the optimal degree $k^*$ is derived for random regular graphs of any size (Supplementary Section 3). Furthermore, in the limit $N\rightarrow
+\infty$, the optimal degree becomes $\sqrt{\frac{Rs}{\left(1-\sum_{i=1}^{n}p_{i}^{2}\right)c}+1}-1$. Since the degree must be an integer in random regular graphs, we have
\begin{equation}
k^{*} \in \left\{ \left\lfloor \sqrt{\frac{Rs}{\left(1-\sum_{i=1}^{n}p_{i}^{2}\right)c}+1}-1 \right\rfloor, \left\lceil \sqrt{\frac{Rs}{\left(1-\sum_{i=1}^{n}p_{i}^{2}\right)c}+1}-1 \right\rceil \right\},
\label{optimal_k}
\end{equation}
where $\lfloor \cdot \rfloor$ and $\lceil \cdot \rceil$ denote the floor and ceiling functions, respectively. The value of $k^*$ is chosen as the one that yields the lower average individual while satisfying the constraints of a regular graph, i.e., $k^{*}\geq 2$. Accordingly, we can see that the optimal $k^{*}$ increases with the baseline reward, but decreases with the task complexity. This means that sparse networks are more likely to promote collective performance for more complex tasks. 
\begin{figure*}[!h]
\centering
\includegraphics[width=16cm]{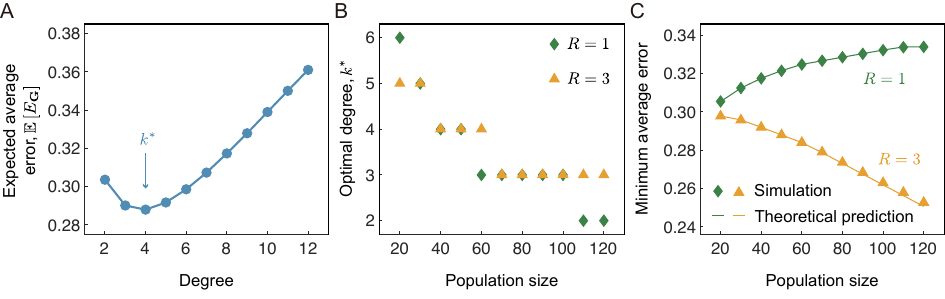}
\caption{\textbf{Optimal degree for collective performance.} (A) Expected average error as a function of degree in random regular graphs. Our theoretical analysis identifies the optimal degree $k^{*}$ that minimizes the average estimation error, thereby maximizing collective performance.
(B) Impact of network size on the optimal degree. The optimal degree $k^{*}$ decreases as the population size increases, and this result is robust across different baseline rewards $R$. This contrasts with the results in the full-sharing scenario, where the optimal degree $k=N-1$ increases with population size. These results underscore the critical role of strategy evolution in shaping collective performance.
(C) Effect of baseline reward and population size on optimal collective performance. For a fixed population size, the best collective performance is achieved in random regular networks with the optimal degree $k^{*}$. Our results show that when the baseline reward is low, increasing the population size leads to higher estimation errors at the corresponding optimal average degree, thereby undermining collective performance. In contrast, a high baseline reward counteracts this effect, enabling larger populations to achieve 
a better collective performance. Parameters: $N=50$ (A), $R=3$ (A), $\theta=0.002$, $n=2$, $p_1=p_2=\frac{1}{2}$, and $s_i=c_i=1$ for $i=1,\cdots,N$.}
\label{Optimal_degree}
\end{figure*}

Our theoretical predictions, validated by numerical simulations (Fig.~3A), accurately identify the optimal degree $k^*$ for collective performance in random regular graphs. We further find that the relationship between the average degree $\langle k \rangle$ and the average individual error follows a consistent pattern, 
similar to that observed in random regular graphs, 
in other network models, including Watts–Strogatz small-world (SW) networks\cite{Watts1998}, Erd\H{o}s--R\'enyi random (ER) networks\cite{erdos1960evolution}, and Barab\'asi--Albert scale-free (SF) networks\cite{Albert2002,Barabasi1999} (Supplementary Fig.~S5). Specifically, the average degree $\langle k \rangle$ plays a critical role in mediating the trade-off between performance under full sharing and the evolution of sharing behavior, thereby shaping the overall group performance. Notably, under the same parameter settings as in Fig.~3A, the optimal average degree for maximizing collective performance in the three network models considered is also close to $4$. This result demonstrates that the optimal degree $k^*$ derived from random regular graphs can serve as a reliable and efficient approximation for the optimal average degree in more general complex networks.

Furthermore, while Eq.~(\ref{optimal_k}) clearly captures the influence of each parameter on the optimal degree in the limit of $N \rightarrow +\infty$, it does not account for the effects of population size. To address this, we further investigate how population size influences the optimal degree. 
The analytical expression for the optimal degree at arbitrary population sizes has been derived in Supplementary Section 3.
An unexpected trend emerges on how collective performance depends on the population size.
In particular, our results show that, as population size increases, the optimal degree decreases regardless of the baseline reward (Fig.~3B), suggesting that a lower degree may be more favorable for collective performance in larger populations. However, even when the optimal degree is the same across different baseline rewards, the corresponding minimum estimation errors can differ (Fig.~3C). In particular, a large baseline reward can markedly enhance the optimal collective performance by reducing the minimum collective estimation error. Moreover, for large baseline rewards, increasing population size further improves collective performance at the optimal degree, whereas for small baseline rewards, larger population size increases the minimum estimation error.

\subsection{Collective performance under heterogeneous allocations}

So far, we have proposed an analytical framework to study collective performance on networks and revealed the critical role of networks in shaping collective behavior. Our previous analysis focused on the uniform case where all individuals draw the same number of samples and incur identical costs for sharing. Indeed, our model can be extended, beyond the uniform case, to heterogeneous cases, in which the number of samples taken by each individual varies. Namely, we consider two typical resource allocation patterns (Fig.~4A), in which the number of samples an individual draws is respectively 
proportional or inversely proportional to its degree, while 
the total number of samples available to the population, $S$, remains fixed. 
To better compare the two heterogeneous cases to the uniform one, we further assume that the cost for a sharer $i$ with $s_i$ samples, $c_i$, scales proportionally with $s_i$ (e.g., $c_i = s_i$).
Although strictly following uniform, proportional, or inversely proportional allocations may result in non-integer numbers of samples per individual, we can apply the largest remainder method to obtain an approximate allocation for each pattern\cite{balinski2001fair} (see Methods).  
\begin{figure*}[!h]
\centering
\includegraphics[width=16cm]{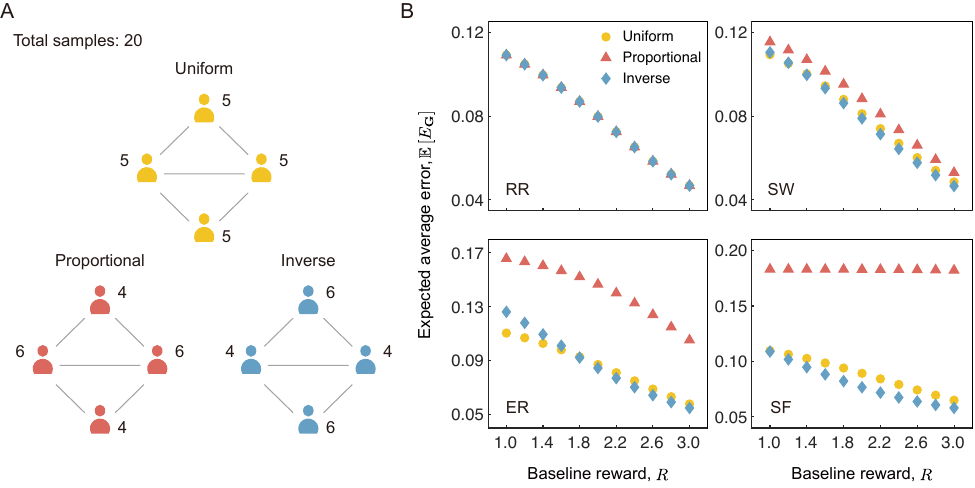}
\caption{\textbf{Effects of sample allocations on collective performance.} (A) Illustration of the three typical sampling allocation schemes considered: uniform (equal number of samples to each individual), proportional (number of samples proportional to an individual’s degree), and inverse (number of samples inversely proportional to an individual’s degree). Each individual is labeled with the number of samples assigned, with the total number of samples $S$ fixed at $20$. 
(B) Average individual error as a function of the baseline reward $R$ for the three sampling patterns, evaluated on four network topologies with average degree $\langle k\rangle = 4$: random regular (RR)\cite{newman2010networks}, Watts--Strogatz small-world (SW)\cite{Watts1998}, Erd\H{o}s--R\'enyi (ER)\cite{erdos1960evolution}, and Barab\'asi--Albert scale-free (SF)\cite{Barabasi1999,Albert2002} graphs. Each symbol represents results from $10^8$ independent simulation realizations (see Methods). Parameters: $N=100$, $S=20$ (A) or $400$ (B), $\theta=0.002$, $n=2$, and $p_1=p_2=\frac{1}{2}$.}
\label{Sample_allocation}
\end{figure*}
We show that, in the absence of strategy evolution, an allocation inversely proportional to node degree cannot be the optimal one among the three allocations considered, either in the full-sharing or in the no-sharing case. In some cases, it even results in the highest average individual error, corresponding to the lowest collective performance (Supplementary Fig.~S6). On the other hand, in the full-sharing state, an allocation proportional to node degree can intuitively enhance collective performance by allowing high-degree nodes to transmit more information (Supplementary Fig.~S6). However, when the evolutionary dynamics of sharing versus non-sharing behaviors are taken into account, the inversely proportional allocation outperforms the two others, achieving the highest collective performance under specific baseline rewards (Fig.~4B). In contrast, the proportional allocation performs worst, yielding the highest average individual error. These findings are consistent across all the different network structures considered in Fig.~4B, offering key insights for the application of our model. Specifically, we reveal a surprising result: under social interactions and certain baseline rewards, collective performance can be improved by allocating more information to lower-degree nodes and less to higher-degree nodes. This is because, although allocating more information to low-degree nodes makes information sharing less effective, it lowers the overall cost of sharing for individuals with many samples. Conversely, allocating more information to high-degree nodes can impede the evolution of sharing behavior, as these nodes incur greater costs for sharing due to their numerous neighbors. Thus, an allocation of information that is inversely proportional to degree may be more favorable for collective performance.
\begin{figure*}[!h]
\centering
\includegraphics[width=16.5cm]{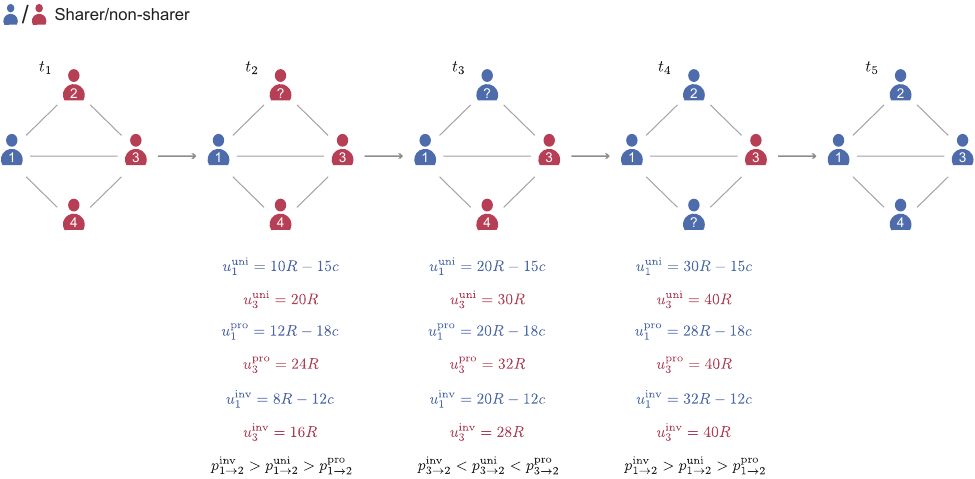}
\caption{\textbf{Intuition for why the inversely proportional allocation enhances collective performance.} We consider an evolutionary process starting from a single sharer at node $1$ in a population of non-sharers (time $t_1$). After social interaction and estimation, the sharer’s total payoff is $u_1^{\text{uni}}$ under uniform sample allocation, while the total payoff of the non-sharer at node $3$ is $u_3^{\text{uni}}$. Similarly, we denote payoffs under degree-proportional and degree-inversely-proportional allocations as $u_1^{\text{pro}}$, $u_3^{\text{pro}}$, and $u_1^{\text{inv}}$, $u_3^{\text{inv}}$, respectively. We then show that the probability of the sharer at node $1$ to invade node $2$ (time $t_2$) satisfies $p_{1\rightarrow}^{\text{inv}}>p_{1\rightarrow 2}^{\text{uni}}>p_{1\rightarrow 2}^{\text{pro}}$, indicating that the spread of sharing behavior from node $1$ to node $2$ is favored when the sample allocation is inversely proportional to node degree. Moreover, after node $2$ adopts the sharer strategy (time $t_3$), it is more likely to resist invasion by non-sharers when the sample allocation is inversely proportional to degree because $p_{3\rightarrow 2}^{\text{inv}}<p_{3\rightarrow 2}^{\text{uni}}<p_{3\rightarrow 2}^{\text{pro}}$. Similar results exist for the imitation process of the individual at node $4$, as node $4$ is symmetric to node $2$. Finally, once individuals at nodes $1$ to $3$ have adopted the sharing strategy, sharers at nodes $4$ (or $2$) are more likely to maintain their sharing behavior when the sample allocation is inversely proportional to degree. Overall, a sample allocation that is inversely proportional to degree increases the likelihood of the population reaching the all-sharer state (time $t_5$), thereby enhancing collective performance. Parameters: $n=2$, $p_1=p_2=\frac{1}{2}$, and $S=20$. The specific sample allocation numbers for three different schemes are the same as those in Fig.~4A.}
\label{Intuition}
\end{figure*}

To further understand how a sample allocation inversely proportional to degree can favor collective performance, we analyze the evolutionary dynamics of sharing behavior in a structured population under the three allocation schemes. Specifically, we start the system with a single sharer in the network, and all other individuals initially adopting a non-sharing strategy\cite{Sheng2024} (Fig.~5). We find that, among the three sample allocation patterns, allocation inversely proportional to node degree most effectively increases the probability that a sharer will invade a neighboring non-sharer, thereby enhancing the spread of the sharing strategy. Conversely, when sharers face potential invasion by non-sharers, an allocation inversely proportional to the degree provides the greatest resistance to such invasions, leading to a higher evolutionary advantage to sharers. Across all stages of the trajectory toward fixation in the toy example of Fig.~5, inversely proportional allocation promotes both the spread and persistence of sharing behavior relative to uniform or degree-proportional allocations. As a result, this pattern favors the evolution of sharing behavior, leading to a persistent all-sharer state and improving overall collective performance.

\section{Discussion}

Collective intelligence is influenced by how the individuals of a group communicate and cooperate. In particular, information sharing is crucial for improving the performance in a collective estimation task\cite{Centola2023,Farrell2011,Becker2017}. However, an individual usually incurs a cost when choosing to cooperate and share information to benefit a neighbor. Consequently, a rational individual is not willing to share information. 
The dilemma of whether cooperating or not is thus an important factor that is entangled with the collective estimation task\cite{Tchernichovski2023}.

In this study, we focus on a collective estimation task in which individuals need to infer the distribution of ball colors in a box. The individuals are arranged in a network and can choose to share information through their connections. The system is modeled as an evolutionary game that integrates the dynamics of strategic information-sharing decisions into the collective estimation process.
We concentrate on how such game interactions influence the collective estimation performance across different network structures.

Our results demonstrate that the collective estimation performance depends significantly on the network. 
Interestingly, we find that the optimal network results from a trade-off between average estimation performance and sharing rate.
In other words, the optimal network is neither the one producing the minimal average individual error under full sharing, nor the one with the highest sharing rate when strategy evolution is considered.
The optimal network ranks instead in the middle range for both these two quantities. 
%
%
Such a surprising result is due to the multifaceted effects of the network on collective performance when strategy evolution is considered.
%
%
It also highlights that we should not only focus on the evolutionary outcomes of collective behavior, but also, importantly, on the quantities capturing the essence of collective estimation performance\cite{Tchernichovski2023}. From this perspective, our proposed framework provides an effective mean to investigate analytically the emergence of collective performance with strategy evolution on various networks. 
%
%
We found that the average degree of a network plays an important role in the final outcome of collective estimation performance. In particular, there exists an intermediate average degree maximizing the collective performance. Through our 
framework we can identify analytically the optimal average degree in random regular graphs. Similar findings are also observed on other types of network by numerical simulations. 
Indeed, studying how network properties influences the emergence of collective intelligence is an important research direction\cite{Becker2017,Centola2022,Centola2023,Baumann2024}. To the best of our knowledge, these findings are not reported in previous works. 
In addition, finding the network that best promotes the emergence of collective intelligence is also very important for practical applications. Identifying such structure through numerical calculations or simulations is challenging especially when the network size are large. Our analytical results thus provide an efficient approach for exploring network structures, thereby offering practical guidance for the design and optimization of real-world networks by adjusting network parameters such as the average degree.


Another important implication of our framework is that it provides an evolutionary explanation for how reward shapes collective performance. In our model, individuals receive rewards based on their estimation accuracy, and strategies evolve according to these payoffs. In particular, the baseline reward modulates the selective advantage of cooperation, influencing the formation of clusters of sharers on networks that resist exploitation by non-sharers. The resulting impact on collective performance is non-trivial and cannot be inferred from intuition alone. While rewards have been employed in previous studies as basic incentives \cite{Lorenz2011,Tchernichovski2023}, our work reveals their mechanistic role in driving strategy evolution. Crucially, this theoretical analysis enables quantitative prediction: for any given reward magnitude, our framework accurately forecasts the expected collective performance (Supplementary Figs.~S3 and S4).

Furthermore, our framework allows us to go beyond uniform sampling and to compare different heterogeneous allocation patterns. Counterintuitively, among the allocation schemes considered, 
we find that degree-proportional allocation leads to the largest average individual estimation error on several non-homogeneous networks, whereas inversely proportional allocation yields the smallest error under certain baseline rewards. This indicates that when strategy evolution is taken into account, allocating limited sampling resources preferentially to high-degree individuals can undermine collective performance, while favoring low-degree nodes may instead enhance collective estimation.
More generally, our results reveal a non-trivial interplay between sampling allocation and evolutionary dynamics, providing guidance for the efficient use of limited sampling resources in networked populations. While we focus here on three representative allocation schemes, an important open question is whether more general or adaptive allocation strategies can further improve collective performance, and what constitutes the optimal allocation rule within an evolutionary framework.

Our framework can be extended in different directions. First of all, in approximating the stationary distribution of sharing and non-sharing states, we employed fixation probabilities as an intermediate step\cite{Wang2022,Hauert2007}. Our theoretical analyses accurately capture fixation probabilities under weak selection, whereas a general analysis applicable to arbitrary selection strengths is NP-hard\cite{Ibsen-Jensen2015}. Our numerical simulations under stronger selection reveal rich and complex dynamics (Supplementary Fig.~S7), so this is a promising research direction
for further investigations.  
%
Secondly, our model mainly focuses on pairwise social interactions. In reality, social interactions often occur in more complex forms, and group of individuals can be better represented through 
higher-order networks\cite{Wang2024,Civilini2021,FerrazdeArruda2020,Majhi2022,Civilini2024,Alvarez-Rodriguez2021,Sheng2024,FerrazdeArruda2023,Battiston2020,Battiston2021,Liu2023}. 
Extending our framework to incorporate such higher-order interactions would not only broaden its applicability but also provide deeper insights into the mechanisms for collective behavior.
%
Furthermore, while information sharing can enhance collective performance when evaluated by the average individual error, it may simultaneously reduce group diversity, as individual estimates become highly correlated after information aggregation, thereby potentially undermining the wisdom of crowds\cite{Lorenz2011,Centola2022}. Nevertheless, the combined effects of network structure and social interactions on group diversity are not yet well understood. Our theoretical framework can be naturally extended to investigate group diversity in estimation tasks, as well as in other types of collective problems\cite{Wang2010a,Iacopini2018,Iacopini2020}. Exploring these directions will provide deeper insights into the underlying mechanisms for the emergence of collective intelligence.

\section{Methods}
Here, we briefly present our analytical framework for studying collective performance on networks, along with a description of our numerical simulations. Detailed derivations are provided in the Supplementary Information.
\subsection{Analysis of average individual error}
According to Eq.~(\ref{eqs:cet2}), the average estimation error is profoundly influenced by the fixation probability of a sharer invading a non-sharing population, $\rho_S$, as well as the fixation probability of a non-sharer invading a population of sharers, $\rho_W$\cite{Ohtsuki2006,Allen2017}. These two quantities determine the stationary distribution between the fully sharing and fully non-sharing states\cite{Wang2022,Hauert2007}. According to our analytical framework, we can provide an approximation for $\rho_S$, which is (see Supplementary Section 2)
\begin{equation}
\rho_S \approx \frac{1}{N}+\frac{\theta}{N}\left( \sum_{j=1}^N \sum_{l=1}^{N} \pi_j \sum_{m=1}^N \left(\left(p_{j m}^{\left(2\right)}-p_{j m}^{\left(0\right)}\right)\frac{R w_{lm} s_l }{1 - \sum_{\alpha=1}^{n} p_{\alpha}^2}-\left(p_{j l}^{\left(2\right)}-p_{j l}^{\left(0\right)}\right)w_{ml}c_l\right)\eta_{jl}\right),
\end{equation}
where $\pi_j=w_j/\sum_{l=1}^N w_l$ is the reproductive value of node $j$, $p_{ij}^{\left(g\right)}$ is the probability of a $g$-step random walk from $i$ to $j$, and $\eta_{jl}$ for $j,l=1,\cdots,N$ are coefficients determined by the population structure, obtained by solving a system of linear equations (see Supplementary Section 2). 

Similarly, for $\rho_W$, we have
\begin{equation}
\rho_W=\frac{1}{N}-\frac{\theta}{N}\left( \sum_{j=1}^N \sum_{l=1}^{N} \pi_j \sum_{m=1}^N \left(\left(p_{j m}^{\left(2\right)}-p_{j m}^{\left(0\right)}\right)\frac{R w_{lm} s_l }{1 - \sum_{\alpha=1}^{n} p_{\alpha}^2}-\left(p_{j l}^{\left(2\right)}-p_{j l}^{\left(0\right)}\right)w_{ml}c_l\right)\eta_{jl}\right).
\label{fix_u_1}
\end{equation}
Therefore, we have $\rho_S+\rho_W\approx 2/N$. According to Eq.~(\ref{eqs:cet2}), the average individual error is then given by
\begin{equation}
\begin{aligned}
\mathbb{E}\left[E_{\mathbf{G}}\right]&=\frac{1-\sum_{\alpha=1}^{n}p_{\alpha}^{2}}{2}\sum_{i=1}^{N}\left(\frac{\rho_S}{s_i+\sum_{l=1}^{N}w_{li}s_l}+\frac{\rho_W}{s_i}\right)\\
&=\frac{\left(1-\sum_{\alpha=1}^{n}p_{\alpha}^{2}\right)\rho_S}{2}\sum_{i=1}^{N}\left(\frac{1}{s_i+\sum_{l=1}^{N}w_{li}s_l}-\frac{1}{s_i}\right)+\sum_{i=1}^{N}\frac{1-\sum_{\alpha=1}^{n}p_{i}^{2}}{Ns_i}.
\end{aligned}
\end{equation}
In random regular graphs with degree $k$\cite{newman2010networks,Ohtsuki2006}, and under the uniform scenario where $s_i=s$ and $c_i=c$ for $i=1,\cdots,N$, the fixation probability of the sharing behavior simplifies to (see Supplementary Section 2)
\begin{equation}
\rho_S \approx \frac{1}{N}+\frac{\theta k}{2N}\left(\frac{Rs}{1-\sum_{i=1}^n p_n^2} \left(\frac{N}{k}-2\right)-c\left(N-2\right)\right).
\label{fix_regular}
\end{equation}
And the corresponding expected average individual error is 
\begin{equation}
\mathbb{E}\left[E_{\mathbf{G}}\right]=\frac{\left(1-\sum_{\alpha=1}^{n}p_{\alpha}^{2}\right)\left(2\left(k+1\right)-Nk\rho_s\right)}{2\left(k+1\right)s}. 
\label{eqs:cevt2}
\end{equation}
Substituting Eq.~(\ref{fix_regular}) into Eq.~(\ref{eqs:cevt2}), the expected average individual error is given by
\begin{equation}
  \mathbb{E}\left[E_{\mathbf{G}}\right]  =\frac{1-\sum_{\alpha=1}^{n}p_{\alpha}^{2}}{s}\left(1-\frac{k}{2\left(k+1\right)}\left(1+k\theta\left( \frac{R s\left(N-2k\right) }{2k\left(1 - \sum_{\alpha=1}^{n} p_{\alpha}^2\right)}-\frac{c\left(N-2\right)}{2}\right)\right)\right).
  \label{eqs:cevt2_1}
\end{equation}
Accordingly, we can determine the optimal degree $k$ that most effectively promotes collective behavior based on Eq.~(\ref{eqs:cevt2_1}) (see Supplementary Section 3).


\subsection{Numerical simulations}
In our investigation, we primarily simulate the fixation probabilities, $\rho_S$ and $\rho_W$, to approximate the stationary distribution measuring the average fraction of time that the population spends in each homogeneous state, as given by Eq.~(\ref{sta}). To simulate $\rho_S$, in each independent run, a randomly selected individual is assigned to share while all others withhold information. We then record the number of samples obtained by each individual during interactions and calculate both the expected rewards and costs incurred. Subsequently, a randomly chosen individual updates its strategy (sharing or not) by imitating a neighbor’s strategy with a probability proportional to the neighbor’s fitness. The population then evolves to the next time step, with social interaction and strategy updating repeated. The maximum number of time steps is set to $10^6$, and simulations that do not reach fixation within this limit are considered unsuccessful. We perform $10^8$ independent simulation runs, calculating the fixation probability $\rho_S$ as the fraction of runs that end in the full-sharing state. The fixation probability $\rho_W$ is obtained through a similar simulation procedure. 

Moreover, to estimate the expected average individual error in each homogeneous state (i.e., full-sharing and non-sharing), we perform $10^6$ independent simulation runs. In each simulation, individual $i$ draws $s_i$ samples from the box for guessing the color distribution $\mathbf{p}$. The total number of samples available to $i$ is $s_i + \sum_{l=1}^{N} w_{li} s_l$ in the full-sharing state or $s_i$ in the non-sharing state. The frequency of each color within the total available samples of individual $i$ is taken as its final estimate. Then, we can compute the average individual error for this simulation run. Finally, we take the mean of these simulation results as the expected average individual error. This process is applied to the two homogeneous states, respectively, to obtain the corresponding expected average error for each state. 

To explore the effects of resource allocation schemes on collective estimation performance, we apply the largest remainder algorithm\cite{balinski2001fair}. Specifically, for a fixed total amount of resources (i.e., total samples), the allocation proceeds as follows. First, we calculate the initial allocation for each individual as the product of the total resources and their assigned proportion. Next, we allocate to each individual the integer part of this product, ensuring that the total initial allocation does not exceed the available resources. Third, we compute the fractional remainder of resources for each individual. The remaining unallocated resources are then distributed sequentially to the individuals with the largest remainders until all resources are assigned. For individuals with identical remainders, the allocation order is determined randomly.

\clearpage

\section*{References}
\bibliography{reference}
\end{document}


\newcommand\bxi{\boldsymbol{\xi}}
\maketitle
\begin{enumerate}
  \item  Center for Systems and Control, College of Engineering, Peking University, Beijing 100871, China
  \item School of Mathematical Sciences, Queen Mary University of London, London E1 4NS, United Kingdom
  \item Sorbonne Universit\'e, Paris Brain Institute (ICM), CNRS UMR7225, INRIA Paris, INSERM U1127, H\^opital de la Piti\'e Salp\^etri\`ere, AP-HP, Paris 75013, France
  \item Division of Decision and Control Systems, School of Electrical Engineering and Computer Science, KTH Royal Institute of Technology, Stockholm 10044, Sweden
  \item Department of Mathematics, KTH Royal Institute of Technology, Stockholm 10044, Sweden
  \item School of Mathematical Sciences, University of Electronic Science and Technology of China, Chengdu 611731, China
  \item Center for Multi-Agent Research, Institute for Artificial Intelligence, Peking University, Beijing 100871, China
  \item Dipartimento di Fisica ed Astronomia, Universit\`a di Catania and INFN, Catania I-95123, Italy
  \item Complexity Science Hub Vienna, Vienna A-1080, Austria
   \item[] Corresponding authors: xiaojiechen@uestc.edu.cn, longwang@pku.edu.cn, \\ v.latora@qmul.ac.uk
\end{enumerate}

\newpage

\tableofcontents

\newpage

The Supplementary Information is organized as follows. 
In Section 1, we present a detailed mathematical analysis for calculating the expected average individual error. In Section 2, we develop a theoretical framework to approximate the stationary distribution measuring the average fraction of time that the population spends in the full-sharing and non-sharing states, thereby providing an analytical approach for evaluating the expected average individual error with strategy evolution.
In Section 3, we determine the optimal degree $k^{*}$ of random regular graphs for maximizing the collective estimation performance, unveiling the effects of network properties for promoting collective intelligence.

\section{Expected average individual error}
As described in the main text, we use the average individual error to quantify the group performance, which captures the overall accuracy of individual estimates. We next provide a detailed derivation of the expected average individual error on networks.

\subsection{Expected average individual error in a single round}
For an individual $i$ with $S_i$ samples, obtained through both individual draws and social interactions in a single round, the corresponding estimate is
\begin{equation}
\mathbf{{p}}^{\left(i\right)} = \left\{{p}_1^{\left(i\right)}, {p}_2^{\left(i\right)}, \dots, {p}_n^{\left(i\right)}\right\},
\end{equation}
where
\begin{equation}
{p}_\alpha^{\left(i\right)} = \frac{X_\alpha^{\left(i\right)}}{S_i}.
\end{equation}
Here, $X_\alpha^{\left(i\right)}$ denotes the observation number of balls with the $\alpha$th color among the $S_i$ samples obtained by individual $i$.

Accordingly, the average individual error in a population with size $N$ is given by \cite{Jayles2017,Centola2022}
\begin{equation}
 \text{Average individual error}=\frac{1}{N}\sum_{i=1}^{N}E_i=\frac{1}{N}\sum_{i=1}^{N}\mathbb{E}\left[\|\mathbf{{p}}^{\left(i\right)} - \mathbf{p}\|^2\right] ,
\end{equation}
where $E_i$ is the estimation error of individual $i$ and $\mathbf{p} = \left\{p_1, p_2, \dots, p_n\right\}$ is the true color distribution. Here, $0 < p_{\alpha} < 1$ for $\alpha = 1, 2, \dots, n$ and $\sum_{\alpha=1}^{n} p_{\alpha} = 1$.

We next provide an analytical derivation of the expected average individual error. We know that for individual $i$, the expected estimation error is
\begin{equation}
\mathbb{E}\left[\|\mathbf{{p}}^{\left(i\right)} - \mathbf{p}\|^2\right] 
= \mathbb{E}\left[\sum_{\alpha=1}^n \left({p}_\alpha^{\left(i\right)} - p_\alpha \right)^2\right] 
= \sum_{\alpha=1}^n \mathbb{E}\left[\left({p}_\alpha^{\left(i\right)} - p_\alpha \right)^2\right].
\label{expect_i}
\end{equation}
Furthermore, for each $\left({p}_{\alpha}^{\left(i\right)} - p_{\alpha} \right)^2$, we have
\begin{equation}
\mathbb{E}\left[\left({p}_{\alpha}^{\left(i\right)} - p_{\alpha}\right)^2\right] 
= \mathbb{E}\left[({p}_{\alpha}^{\left(i\right)})^2 - 2 {p}_{\alpha}^{\left(i\right)} p_{\alpha} + p_{\alpha}^2\right] 
= \mathbb{E}\left[({p}_{\alpha}^{\left(i\right)})^2\right] - 2 p_{\alpha} \mathbb{E}\left[{p}_{\alpha}^{\left(i\right)}\right] + p_{\alpha}^2.
\label{expect_single}
\end{equation}
Since $\mathbb{E}\left[{p}_{\alpha}^{\left(i\right)}\right] = p_{\alpha}$, Eq.~(\ref{expect_single}) simplifies to
\begin{equation}
\mathbb{E}\left[\left({p}_{\alpha}^{\left(i\right)} - p_{\alpha}\right)^2\right] = \mathbb{E}\left[\left({p}_{\alpha}^{\left(i\right)}\right)^2\right] - p_{\alpha}^2 = \text{Var}\left({p}_{\alpha}^{\left(i\right)}\right),
\label{expect_single_sim}
\end{equation}
where $\text{Var}\left({p}_{\alpha}^{\left(i\right)}\right)$ is the variance of ${p}_{\alpha}^{\left(i\right)}$, which is given by
\begin{equation}
\text{Var}\left({p}_{\alpha}^{\left(i\right)}\right)=\text{Var}\left(\frac{X_{\alpha}^{\left(i\right)}}{S_i}\right)=\frac{\text{Var}\left({X_{\alpha}^{\left(i\right)}}\right)}{S_i ^{2}}.
\label{var_m}
\end{equation}
Indeed, $X_{\alpha}^{\left(i\right)}$ follows a binomial distribution (as a marginal of the multinomial distribution), with variance
\begin{equation}
\text{Var}(X_{\alpha}^{\left(i\right)}) = S_i \, p_{\alpha} \left(1 - p_{\alpha}\right).
\label{var_m_specific}
\end{equation}
Therefore, based on Eqs.~(\ref{expect_i})-(\ref{var_m_specific}), we obtain the specific form of the expected estimation error for individual $i$
\begin{equation}
\mathbb{E}\left[\|\hat{\mathbf p}^{\left(i\right)} - \mathbf p\|^2\right] =\sum_{\alpha=1}^n  \frac{p_\alpha \left(1-p_\alpha\right)}{S_i}=\frac{1-\sum_{\alpha=1}^n p_{\alpha} ^{2}}{S_i}.  
\end{equation}
Here, $1-\sum_{\alpha=1}^n p_{\alpha} ^{2}$ quantifies the task complexity, with a more uneven distribution or a larger number of color categories corresponding to higher complexity.

Therefore the expected average individual error is obtained by averaging the expected estimation errors across all individuals, which is given by
\begin{equation}
 \mathbb{E}\left[\text{Average individual error}\right]=\frac{1}{N}\sum_{i=1}^{N}\frac{1-\sum_{\alpha=1}^n p_{\alpha} ^{2}}{S_i}.  
\end{equation}
\subsection{Expected average individual error under social interactions}
So far, we have derived the expected average individual error based on the samples available to each participant in a single round. Notably, when social interactions are incorporated into the dynamical process, the samples available to each individual may vary depending on the behavior of their neighbors, leading to evolving estimation errors. Accordingly, the expected average individual error is defined by the limit of average individual error per round over an extremely long time interval
\begin{equation}
    \mathbb{E}\left[\text{Average individual error}\right]=\frac{1}{T}\sum_{t=1}^{T}\sum_{i=1}^{N}\frac{1-\sum_{\alpha=1}^{n}p_{\alpha}^{2}}{N S_i\left(t\right)},
\end{equation}
where $S_i\left(t\right)$ is the number of samples obtained by individual $i$ at time step $t$.

Furthermore, as discussed in the main text, the underlying population structure for social interactions is modeled by a network $\mathbf{G}=\left[w_{ij}\right]_{i,j=1}^{N}$, with $w_{li}=1$ if $l$ and $i$ are neighbors and $0$ otherwise. Let the population state at time $t$ be $\mathbf{x} = \{x_1, x_2, \dots, x_N\}$, where $x_i = 1$ if individual $i$ chooses to share information and $x_i = 0$ otherwise. The total number of samples available to individual $i$ at this time is then given by
\begin{equation}
 S_i\left(t\right)=s_i+ \sum_{l=1}^{N} w_{li} s_l x_l, 
\end{equation}
where $s_i$ is the number of samples drawn by individual $i$, and $\sum_{l=1}^{N} w_{li} s_l x_l$ represents the number of samples that individual $i$ receives from its neighbors. Accordingly, the expected average individual error on network $\mathbf{G}$, $\mathbb{E}\left[E_{\mathbf{G}}\right]$, is written as
\begin{equation}
 \mathbb{E}\left[E_{\mathbf{G}}\right]=\frac{1}{T}\sum_{t=1}^{T}\sum_{i=1}^{N}\frac{1-\sum_{\alpha=1}^{n}p_{\alpha}^{2}}{N \left(s_i+ \sum_{l=1}^{N} w_{li} s_l x_l\right)}.
 \label{Exp_Gr_State}
\end{equation}
Since our model incorporates a rare exploration process, the population evolves toward either a full-sharing or a full-non-sharing state. In the full-sharing state, the total number of samples available to individual $i$ is
\begin{equation}
S_i = s_i + \sum_{l=1}^{N} w_{li} s_l.
\end{equation}
In contrast, in the full-non-sharing state, $S_i$ reduces to
\begin{equation}
S_i = s_i.
\end{equation}
Thus, the expected average individual error in the full-sharing ($\mathbb{E}\left[E_{\mathbf{G}}^S\right]$) and full-non-sharing ($\mathbb{E}\left[E_{\mathbf{G}}^W\right]$) states are respectively given by
\begin{equation}
\begin{cases}
  \mathbb{E}\left[E_{\mathbf{G}}^S\right]= \sum_{i=1}^{N}\frac{1-\sum_{\alpha=1}^{n}p_{\alpha}^{2}}{N \left(s_i + \sum_{l=1}^{N} w_{li} s_l\right)},\\
  \mathbb{E}\left[E_{\mathbf{G}}^W\right]=  \sum_{i=1}^{N}\frac{1-\sum_{\alpha=1}^{n}p_{\alpha}^{2}}{N s_i}.
\end{cases}
\end{equation}
Accordingly, the expected average individual error is mainly determined by the time that the population spends in the full-sharing and full-non-sharing states since the time spent in intermediate states can be neglected under the assumption of rare exploration. Let $t_S$ ($t_W$) be the average fraction of time that the population spends in the full-sharing (full-non-sharing) state, thus we have 
\begin{equation}
\begin{aligned}
\mathbb{E}\left[E_{\mathbf{G}}\right]&=t_S\mathbb{E}\left[E_{\mathbf{G}}^S\right]+t_W\mathbb{E}\left[E_{\mathbf{G}}^W\right] \\
 &=t_S\sum_{i=1}^{N}\frac{1-\sum_{\alpha=1}^{n}p_{\alpha}^{2}}{N \left(s_i + \sum_{l=1}^{N} w_{li} s_l\right)}+t_W\sum_{i=1}^{N}\frac{1-\sum_{\alpha=1}^{n}p_{\alpha}^{2}}{N s_i}.   
 \label{exp_GE}
\end{aligned}
\end{equation}
\section{Approximation of the expected average individual error}
The fraction of time that the population spends in each state can be obtained by computing the stationary distribution of the corresponding Markov chain \cite{Fudenberg2006}, which consists of two homogeneous states: full-sharing and full-non-sharing. To determine this distribution, we first introduce the transition matrix $\mathbf{M}$, which describes the transitions between the full-sharing and full-non-sharing states, and is given by
\begin{equation}
 \mathbf{M} =\left[\begin{array}{cc}
1-\mu \rho_{W}  & \mu \rho_{W} \\
\mu \rho_{S}  & 1-\mu \rho_{S} 
\end{array}\right],  
\end{equation}
where $\mu$ is the exploration (mutation) rate, and $\rho_S$ ($\rho_W$) is the fixation probability of a sharer (non-sharer) invading and taking over a population of non-sharers (sharers). Accordingly, $\mu \rho_S$ ($\mu \rho_W$) corresponds to the probability that a population of sharers (non-sharers) transitions to a population of non-sharers (sharers). Here, $\rho_S$ and $\rho_W$ are defined in the absence of exploration. Under the assumption of rare exploration, these fixation probabilities can be used to characterize the long-term evolutionary dynamics of the population.

Furthermore, the stationary distribution of the population across the two states, $\mathbf{t}=\left[t_S, t_W\right]$, satisfies $\mathbf{tM}=\mathbf{t}$ and $t_S+t_W=1$. Moreover, the specific form of $\mathbf{t}$ is obtained as \cite{Fudenberg2006,Wang2022}
\begin{equation}
 t_S=\frac{\rho_S}{\rho_S +\rho_W},  
 \label{t_S}
\end{equation}
\begin{equation}
 t_W=\frac{\rho_W}{\rho_S +\rho_W}.  
 \label{t_W}
\end{equation}
According to Eqs. (\ref{exp_GE}) and (\ref{t_S})-(\ref{t_W}), the expected average individual error can be approximated using the fixation probabilities
\begin{equation}
\begin{aligned}
\mathbb{E}\left[E_{\mathbf{G}}\right]&=\frac{\rho_S}{\rho_S+\rho_W}\sum_{i=1}^{N}\frac{1-\sum_{\alpha=1}^{n}p_{\alpha}^{2}}{N \left(s_i + \sum_{l=1}^{N} w_{li} s_l\right)}+\frac{\rho_W}{\rho_S+\rho_W}\sum_{i=1}^{N}\frac{1-\sum_{\alpha=1}^{n}p_{\alpha}^{2}}{N s_i}  \\
 &= \frac{\rho_S}{\rho_S+\rho_W}\left(\sum_{i=1}^{N}\frac{1-\sum_{\alpha=1}^{n}p_{\alpha}^{2}}{N \left(s_i + \sum_{l=1}^{N} w_{li} s_l\right)}-\sum_{i=1}^{N}\frac{1-\sum_{\alpha=1}^{n}p_{\alpha}^{2}}{N s_i}\right)+\sum_{i=1}^{N}\frac{1-\sum_{\alpha=1}^{n}p_{\alpha}^{2}}{N s_i}.
 \label{exp_GE_fix}
\end{aligned}
\end{equation}
Next, we respectively calculate the fixation probabilities, $\rho_S$ and $\rho_W$, to obtain an explicit expression for the expected average individual error.

\subsection{Calculation of $\rho_S$}
We begin by deriving the specific form of $\rho_S$. According to Eq.~(\ref{Exp_Gr_State}), the expected estimation error for individual $i$ in a single round is
\begin{equation}
E_i = \frac{1 - \sum_{\alpha=1}^{n} p_{\alpha}^2}{s_i + \sum_{l=1}^{N} w_{li} s_l x_l}.
\end{equation}
Accordingly, as discussed in the main text, the reward received by individual $i$ is inversely proportional to its estimation error, which can be described by the reward function $f\left(E_i\right)$. Let $f\left(E_i\right)=1/E_i$, then the total payoff of individual $i$, $u_i$ (defined as reward minus cost), is given by
\begin{equation}
u_i \left(\mathbf{x}\right) = f\left(E_i\right)-\sum_{l=1}^{N} w_{li} c_i x_i =
\frac{R\left(s_i + \sum_{l=1}^{N} w_{li} s_l x_l\right)}{1 - \sum_{\alpha=1}^{n} p_{\alpha}^2} - \sum_{l=1}^{N} w_{li} c_i x_i.
\label{payoff_function}
\end{equation}
We further define the fitness of individual $i$, determining the outcome of competition among different strategies during the evolutionary process, and is given by
\begin{equation}
F_{i} \left(\mathbf{x}\right) =\exp{\left(\theta u_i \left(\mathbf{x}\right)\right)},
\label{fitness}
\end{equation}
where $\theta$ is the selection strength.

Subsequently, in the strategy updating process, a randomly chosen individual updates its strategy by imitating that of a neighbor, with a probability proportional to the neighbor’s fitness. Thus, the probability that individual $i$ transmits its strategy to individual $j$ is \cite{McAvoy2021}
\begin{equation}
e_{ij} \left(\mathbf{x}\right)=\frac{1}{N}\frac{F_i \left(\mathbf{x}\right)w_{ij}}{\sum_{m=1}^{N}F_m \left(\mathbf{x}\right)w_{mj}}. 
\label{e}
\end{equation}
Differentiating with respect to $\theta$ at $\theta=0$, we have
\begin{equation}
\begin{aligned}
\left.\partial_\theta e_{i j}\left(\mathbf{x}\right)\right|_{\theta=0} &=\frac{1}{N}\frac{u_{i}\left(\mathbf{x}\right)w_{ij}w_j-w_{ij}\sum_{m=1}^{N}u_{m}\left(\mathbf{x}\right)w_{mj}}{w_{j}^{2}} \\&=\frac{p_{j i}^{\left(1\right)}}{N}\left(u_i\left(\mathbf{x}\right)-\sum_{m=1}^N p_{j m}^{\left(1\right)} u_m\left(\mathbf{x}\right)\right)=\frac{p_{j i}^{\left(1\right)}}{N}\left(\sum_{m=1}^N\left(p_{i m}^{\left(0\right)}-p_{j m}^{\left(1\right)}\right) u_m\left(\mathbf{x}\right)\right),
\end{aligned}
\label{diff}
\end{equation}
where $p_{ji}^{\left(g\right)}$ is the probability of a $g$-step random walk from $j$ to $i$.

Substituting Eq.~(\ref{payoff_function}) into (\ref{diff}) gives
\begin{equation}
\begin{aligned}
\left.\partial_\theta e_{i j}\left(\mathbf{x}\right)\right|_{\theta=0} &=\frac{p_{j i}^{\left(1\right)}}{N}\left(\sum_{m=1}^N\left(p_{i m}^{\left(0\right)}-p_{j m}^{\left(1\right)}\right) \left(
\frac{R\left(s_m + \sum_{l=1}^{N} w_{lm} s_l x_l\right)}{1 - \sum_{\alpha=1}^{n} p_{\alpha}^2} - \sum_{l=1}^{N} w_{lm} c_m x_m\right)\right)\\
&=\sum_{I\in \mathcal{L}}d_{I}^{ij}x_{I},
\end{aligned}
\label{diff_1}
\end{equation}
where $\mathcal{L}=\emptyset \cup \left\{1,\cdots,N\right\}$ and $x_{\emptyset}=1$. Therefore, we have
\begin{equation}
    d_{I}^{ij}=\begin{cases} 
\frac{p_{j i}^{\left(1\right)}}{N}\sum_{m=1}^N 
\frac{\left(p_{i m}^{\left(0\right)}-p_{j m}^{\left(1\right)}\right)Rs_m}{1 - \sum_{\alpha=1}^{n} p_{\alpha}^2}  & I=\emptyset, \\ 
\frac{p_{j i}^{\left(1\right)}}{N}\sum_{m=1}^N \left(\left(p_{i m}^{\left(0\right)}-p_{j m}^{\left(1\right)}\right)\frac{R  w_{lm} s_l }{1 - \sum_{\alpha=1}^{n} p_{\alpha}^2}-\left(p_{i l}^{\left(0\right)}-p_{j l}^{\left(1\right)}\right)w_{ml}c_l\right) & I=l\neq \emptyset .
\end{cases}
\label{form_d}
\end{equation}
Furthermore, the fixation probability of a sharing individual to invade and take over a population of non-sharers, starting from a specific initial configuration $\bxi = (\xi_1,...,\xi_N)^\text{T}$ can be approximated as \cite{Allen2017,McAvoy2021}
\begin{equation}
\rho_S(\boldsymbol{\xi})=\widehat{\xi}+\theta\left(\sum_{i=1}^N \pi_i \sum_{j=1}^N \sum_{I \in \mathcal{L}} d_I^{j i}\left(\eta_{\{i\} \cup I}^{\xi}-\eta_{\{j\} \cup I}^{\xi}\right)\right)+O\left(\theta^2\right),
\label{fix_s}
\end{equation}
where $\widehat{\xi}=\sum_{i=1}^N \pi_i \xi_i$ and the terms $\eta_{ij}^{\bxi}$ arise as the unique solution to the following equations
\begin{equation}
\begin{aligned}
\eta_{i j}^{\bxi} & =\frac{N}{2}\left(\widehat{\xi}-\xi_i \xi_j\right)+\frac{1}{2} \sum_{y=1}^N\left(p_{i y}^{(1)} \eta_{y j}^{\bxi}+p_{j y}^{(1)} \eta_{i y}^{\bxi}\right) \quad(i \neq j) ; \\
\eta_{i i}^{\bxi} & =N\left(\widehat{\xi}-\xi_i\right)+\sum_{j=1}^N p_{i j}^{(1)} \eta_{j j}^{\bxi} ; \\
\sum_{i=1}^N \pi_i \eta_{i i}^{\bxi} & =0 .
\end{aligned}
\label{xi_1}
\end{equation}
We focus on the case of uniform initialization, $\mathbf{u}$, in which a single sharer is randomly introduced into a population of non-sharers. In this scenario, the fixation probability of the sharer is given by
\begin{equation}
\begin{aligned}
\rho_S\left(\mathbf{u}\right)&=\frac{1}{N}+\theta\left(\sum_{i=1}^N \pi_i \sum_{j=1}^N \sum_{I \in \mathcal{L}} d_I^{j i}\left(\eta_{\{i\} \cup I}^{\mathbf{u}}-\eta_{\{j\} \cup I}^{\mathbf{u}}\right)\right)+O\left(\theta^2\right)
\end{aligned}.
\label{fix_u}
\end{equation}
Similarly, the terms $\eta^{\mathbf{u}}_{ij}$ satisfy the following equations
\begin{equation}
\begin{aligned}
\eta_{i j}^{\mathbf{u}} & =\frac{1}{2}+\frac{1}{2} \sum_{y=1}^N\left(p_{i y}^{(1)} \eta_{y j}^{\mathbf{u}}+p_{j y}^{(1)} \eta_{i y}^{\mathbf{u}}\right) \quad(i \neq j) ; \\
\eta_{i i}^{\mathbf{u}} & =0 .
\end{aligned}
\label{xi_2}
\end{equation}
According to Eqs.~(\ref{form_d}) and (\ref{fix_u})-(\ref{xi_2}), the fixation probability $\rho_S\left(\mathbf{u}\right)$ is given by
\begin{equation}
\begin{aligned}
\rho_S\left(\mathbf{u}\right)&=\frac{1}{N}+\theta\left(\sum_{i=1}^N \pi_i \sum_{j=1}^N \sum_{l=1}^{N} \frac{p_{i j}^{\left(1\right)}}{N}\sum_{m=1}^N \left(\left(p_{j m}^{\left(0\right)}-p_{i m}^{\left(1\right)}\right)\frac{R w_{lm} s_l }{1 - \sum_{\alpha=1}^{n} p_{\alpha}^2}-\left(p_{j l}^{\left(0\right)}-p_{i l}^{\left(1\right)}\right)w_{ml}c_l\right)\left(\eta_{il}^{\mathbf{u}}-\eta_{jl }^{\mathbf{u}}\right)\right)\\&+O\left(\theta^2\right)\\
&=\frac{1}{N}+\frac{\theta}{N}\left( \sum_{j=1}^N \sum_{l=1}^{N} \pi_j \sum_{m=1}^N \left(\left(p_{j m}^{\left(2\right)}-p_{j m}^{\left(0\right)}\right)\frac{R w_{lm} s_l }{1 - \sum_{\alpha=1}^{n} p_{\alpha}^2}-\left(p_{j l}^{\left(2\right)}-p_{j l}^{\left(0\right)}\right)w_{ml}c_l\right)\eta_{jl}^{\mathbf{u}}\right)+O\left(\theta^2\right).\\
\end{aligned}
\label{fix_u_2}
\end{equation}
Under weak selection where $\theta \ll 1$, we have
\begin{equation}
\rho_S\left(\mathbf{u}\right)\approx\frac{1}{N}+\frac{\theta}{N}\left( \sum_{j=1}^N \sum_{l=1}^{N} \pi_j \sum_{m=1}^N \left(\left(p_{j m}^{\left(2\right)}-p_{j m}^{\left(0\right)}\right)\frac{R w_{lm} s_l }{1 - \sum_{\alpha=1}^{n} p_{\alpha}^2}-\left(p_{j l}^{\left(2\right)}-p_{j l}^{\left(0\right)}\right)w_{ml}c_l\right)\eta_{jl}^{\mathbf{u}}\right).
\label{fix_u_1}
\end{equation}

\subsection{Calculation of $\rho_W$}
Similarly, within our theoretical framework, the fixation probability of a non-sharer invading and taking over the population of sharers can also be derived analytically. 

Specifically, let the population state be $\hat{\mathbf{x}} = \{\hat{x}_1, \hat{x}_2, \dots, \hat{x}_N\}$, where $\hat{x}_i = 1$ if individual $i$ is a non-sharer and $0$ otherwise. According to Eq.~(\ref{payoff_function}), the payoff of individual $i$ is
\begin{equation}
\begin{aligned}
 \hat{u}_i \left(\hat{\mathbf{x}}\right) &= \frac{R\left(s_i + \sum_{l=1}^{N} w_{li} s_l \left(1-\hat{x}_l\right)\right)}{1 - \sum_{\alpha=1}^{n} p_{\alpha}^2} - \sum_{l=1}^{N} w_{li} c_i \left(1-\hat{x}_i\right) \\
 &=\frac{R\left(s_i + \sum_{l=1}^{N} w_{li} s_l\right)}{1 - \sum_{\alpha=1}^{n} p_{\alpha}^2} - \sum_{l=1}^{N} w_{li} c_i -\left(\frac{R \sum_{l=1}^{N} w_{li} s_l \hat{x}_l}{1 - \sum_{\alpha=1}^{n} p_{\alpha}^2} - \sum_{l=1}^{N} w_{li} c_i \hat{x}_i\right).
\end{aligned}
\label{payoff_function_1}
\end{equation}
Furthermore, based on Eq.~(\ref{fitness})-(\ref{form_d}), we have
\begin{equation}
\begin{aligned}
\left.\partial_\theta e_{i j}\left(\hat{\mathbf{x}}\right)\right|_{\theta=0} &=\frac{p_{j i}^{\left(1\right)}}{N}\left(\sum_{m=1}^N\left(p_{i m}^{\left(0\right)}-p_{j m}^{\left(1\right)}\right) \left(\frac{R\left(s_m + \sum_{l=1}^{N} w_{lm} s_l\right)}{1 - \sum_{\alpha=1}^{n} p_{\alpha}^2} - \sum_{l=1}^{N} w_{li} c_i -\frac{R \sum_{l=1}^{N} w_{lm} s_l \hat{x}_l}{1 - \sum_{\alpha=1}^{n} p_{\alpha}^2} + \sum_{l=1}^{N} w_{lm} c_m \hat{x}_m\right)\right)\\
&=\sum_{I\in \mathcal{L}}\hat{d}_{I}^{ij}\hat{x}_{I}
\end{aligned},
\label{diff_2}
\end{equation}
where $\mathcal{L}=\emptyset \cup \left\{1,\cdots,N\right\}$ and $\hat{x}_{\emptyset}=1$. Similar to Eq.~(\ref{form_d}), we further obtain
\begin{equation}
    \hat{d}_{I}^{ij}=\begin{cases} 
\frac{p_{j i}^{\left(1\right)}}{N}\sum_{m=1}^N 
\left(p_{i m}^{\left(0\right)}-p_{j m}^{\left(1\right)}\right) \left(\frac{R\left(s_m + \sum_{l=1}^{N} w_{lm} s_l\right)}{1 - \sum_{\alpha=1}^{n} p_{\alpha}^2} - \sum_{l=1}^{N} w_{li} c_i \right)  & I=\emptyset, \\ 
-\frac{p_{j i}^{\left(1\right)}}{N}\sum_{m=1}^N \left(\left(p_{i m}^{\left(0\right)}-p_{j m}^{\left(1\right)}\right)\frac{R  w_{lm} s_l }{1 - \sum_{\alpha=1}^{n} p_{\alpha}^2}-\left(p_{i l}^{\left(0\right)}-p_{j l}^{\left(1\right)}\right)w_{ml}c_l\right) & I=l\neq \emptyset. 
\end{cases}
\label{form_d_1}
\end{equation}
Accordingly, under the initial configuration $\bxi = (\xi_1,...,\xi_N)^\text{T}$, the fixation probability of a non-sharer invading and taking over a population of sharers, $\rho_W(\boldsymbol{\xi})$, is given by \cite{Allen2017,McAvoy2021}
\begin{equation}
\rho_W(\boldsymbol{\xi})=\widehat{\xi}+\theta\left(\sum_{i=1}^N \pi_i \sum_{j=1}^N \sum_{I \in \mathcal{L}} \hat{d}_I^{j i}\left(\eta_{\{i\} \cup I}^{\xi}-\eta_{\{j\} \cup I}^{\xi}\right)\right)+O\left(\theta^2\right),
\end{equation}
where $\widehat{\xi}$ and the terms $\eta_{ij}^{\bxi}$ are consistent with those in Eq.~(\ref{fix_s}).

Moreover, under uniform initialization $\mathbf{u}$, we have
\begin{equation}
\rho_W\left(\mathbf{u}\right)=\frac{1}{N}+\theta\left(\sum_{i=1}^N \pi_i \sum_{j=1}^N \sum_{I \in \mathcal{L}} \hat{d}_I^{j i}\left(\eta_{\{i\} \cup I}^{\mathbf{u}}-\eta_{\{j\} \cup I}^{\mathbf{u}}\right)\right)+O\left(\theta^2\right),
\label{fix_u_w}
\end{equation}
where the terms $\eta^{\mu}_{ij}$ also satisfy Eq.~(\ref{xi_2}).

According to Eqs.~(\ref{diff_2})-(\ref{form_d_1}) and (\ref{fix_u_w}), we obtain the explicit form of $\rho_W\left(\mathbf{u}\right)$
\begin{equation}
\begin{aligned}
\rho_W\left(\mathbf{u}\right)&=\frac{1}{N}-\theta\left(\sum_{i=1}^N \pi_i \sum_{j=1}^N \sum_{l=1}^{N} \frac{p_{i j}^{\left(1\right)}}{N}\sum_{m=1}^N \left(\left(p_{j m}^{\left(0\right)}-p_{i m}^{\left(1\right)}\right)\frac{R w_{lm} s_l }{1 - \sum_{\alpha=1}^{n} p_{\alpha}^2}-\left(p_{j l}^{\left(0\right)}-p_{i l}^{\left(1\right)}\right)w_{ml}c_l\right)\left(\eta_{il}^{\mathbf{u}}-\eta_{jl }^{\mathbf{u}}\right)\right)\\&+O\left(\theta^2\right)\\
&=\frac{1}{N}-\frac{\theta}{N}\left( \sum_{j=1}^N \sum_{l=1}^{N} \pi_j \sum_{m=1}^N \left(\left(p_{j m}^{\left(2\right)}-p_{j m}^{\left(0\right)}\right)\frac{R w_{lm} s_l }{1 - \sum_{\alpha=1}^{n} p_{\alpha}^2}-\left(p_{j l}^{\left(2\right)}-p_{j l}^{\left(0\right)}\right)w_{ml}c_l\right)\eta_{jl}^{\mathbf{u}}\right)+O\left(\theta^2\right).\\
\end{aligned}
\label{fix_u_w_explicit}
\end{equation}
In addition, under weak selection, we have
\begin{equation}
\rho_W\left(\mathbf{u}\right)\approx\frac{1}{N}-\frac{\theta}{N}\left( \sum_{j=1}^N \sum_{l=1}^{N} \pi_j \sum_{m=1}^N \left(\left(p_{j m}^{\left(2\right)}-p_{j m}^{\left(0\right)}\right)\frac{R w_{lm} s_l }{1 - \sum_{\alpha=1}^{n} p_{\alpha}^2}-\left(p_{j l}^{\left(2\right)}-p_{j l}^{\left(0\right)}\right)w_{ml}c_l\right)\eta_{jl}^{\mathbf{u}}\right).
\label{fix_u_w_1}
\end{equation}

\subsection{Approximation of the expected average individual error via fixation probabilities}
Since we have derived the explicit form of $\rho_S$ and $\rho_W$, the expected average individual error shown in Eq.~(\ref{exp_GE_fix}) can be obtained analytically.
Under uniform initialization, the relationship between $\rho_S\left(\mathbf{u}\right)$ and $\rho_W \left(\mathbf{u}\right)$ (simplified as $\rho_S$ and $\rho_W$) can be approximated based on Eqs.~(\ref{fix_u_1}) and (\ref{fix_u_w_1}), which is given by
\begin{equation}
\rho_S+\rho_W=\frac{2}{N}.
\label{rho}
\end{equation}
Substituting Eqs.~(\ref{fix_u_1}) and (\ref{rho}) into Eq.~(\ref{exp_GE_fix}), we have
\begin{equation}
\begin{aligned}
\mathbb{E}\left[E_{\mathbf{G}}\right]=& \frac{1}{2}\left(1+{\theta}\left( \sum_{j=1}^N \sum_{l=1}^{N} \pi_j \sum_{m=1}^N \left(\left(p_{j m}^{\left(2\right)}-p_{j m}^{\left(0\right)}\right)\frac{R w_{lm} s_l }{1 - \sum_{\alpha=1}^{n} p_{\alpha}^2}-\left(p_{j l}^{\left(2\right)}-p_{j l}^{\left(0\right)}\right)w_{ml}c_l\right)\eta_{jl}^{\mathbf{u}}\right)\right)\\&\left(\sum_{i=1}^{N}\frac{1-\sum_{\alpha=1}^{n}p_{\alpha}^{2}}{N \left(s_i + \sum_{l=1}^{N} w_{li} s_l\right)}-\sum_{i=1}^{N}\frac{1-\sum_{\alpha=1}^{n}p_{\alpha}^{2}}{N s_i}\right)+\sum_{i=1}^{N}\frac{1-\sum_{\alpha=1}^{n}p_{\alpha}^{2}}{N s_i},
 \label{exp_GE_fix_1}
\end{aligned}
\end{equation}
where $\eta_{jl}^{\mathbf{u}}$ follows the definition in Eq.~(\ref{xi_2}) and can be simplified to $\eta_{jl}$.

Taking regular graphs as an example, where each individual has $k$ neighbors, i.e., $w_i = k$ for $i = 1, \dots, N$, we accordingly have
\begin{equation}
\rho_S\approx\frac{1}{N}+\frac{\theta}{N^2}\left( \sum_{j=1}^N \sum_{l=1}^{N} \sum_{m=1}^N \left(\left(p_{j m}^{\left(2\right)}-p_{j m}^{\left(0\right)}\right)\frac{R w_{lm} s_l }{1 - \sum_{\alpha=1}^{n} p_{\alpha}^2}-\left(p_{j l}^{\left(2\right)}-p_{j l}^{\left(0\right)}\right)w_{ml}c_l\right)\eta_{jl}\right).
\label{fix_u_w_r}
\end{equation}
Furthermore, in the homogeneous case where $s_i=s$ and $c_i=c$ for $i=1,\cdots,N$, Eq.~({\ref{fix_u_w_r}}) simplifies to
\begin{equation}
\begin{aligned}
\rho_S&\approx\frac{1}{N}+\frac{\theta}{N^2}\left( \sum_{j=1}^N \sum_{l=1}^{N} \sum_{m=1}^N \left(\left(p_{j m}^{\left(2\right)}-p_{j m}^{\left(0\right)}\right)\frac{R w_{lm} s }{1 - \sum_{\alpha=1}^{n} p_{\alpha}^2}-\left(p_{j l}^{\left(2\right)}-p_{j l}^{\left(0\right)}\right)w_{ml}c\right)\eta_{jl}\right) \\&=\frac{1}{N}+\frac{\theta}{N^2}\left( \sum_{j=1}^N \sum_{l=1}^{N} \sum_{m=1}^N \left(\left(p_{j m}^{\left(2\right)}-p_{j m}^{\left(0\right)}\right)\frac{R kp_{ml}^{\left(1\right)} s }{1 - \sum_{\alpha=1}^{n} p_{\alpha}^2}-\left(p_{j l}^{\left(2\right)}-p_{j l}^{\left(0\right)}\right)kp_{lm}^{\left(1\right)}c\right)\eta_{jl}\right)\\&=\frac{1}{N}+\frac{k\theta}{N^2}\left( \sum_{j=1}^N \sum_{l=1}^{N} \sum_{m=1}^N \left(\left(p_{j m}^{\left(2\right)}-p_{j m}^{\left(0\right)}\right)\frac{R p_{ml}^{\left(1\right)} s }{1 - \sum_{\alpha=1}^{n} p_{\alpha}^2}-\left(p_{j l}^{\left(2\right)}-p_{j l}^{\left(0\right)}\right)p_{lm}^{\left(1\right)}c\right)\eta_{jl}\right)\\&=\frac{1}{N}+\frac{k\theta}{N^2}\left( \sum_{j=1}^N \sum_{l=1}^{N}  \left(\left(p_{j l}^{\left(3\right)}-p_{j l}^{\left(1\right)}\right)\frac{R s }{1 - \sum_{\alpha=1}^{n} p_{\alpha}^2}-\left(p_{j l}^{\left(2\right)}-p_{j l}^{\left(0\right)}\right)c\right)\eta_{jl}\right).
\end{aligned}
\label{fix_u_w_r_1}
\end{equation}
Following the proposed analytical framework \cite{Allen2017,McAvoy2021}, we obtain
\begin{equation}
 \begin{cases}
  \sum_{j=1}^N \sum_{l=1}^{N} \left(p_{j l}^{\left(3\right)}-p_{j l}^{\left(1\right)}\right)\eta_{jl}=\frac{N\left(N-2k\right)}{2k}, \\
  \sum_{j=1}^N \sum_{l=1}^{N} \left(p_{j l}^{\left(2\right)}-p_{j l}^{\left(0\right)}\right)\eta_{jl}=\frac{N^2-2N}{2}.
 \end{cases}   
 \label{sim_eta}
\end{equation}
Substituting Eq.~(\ref{sim_eta}) into Eq.~(\ref{fix_u_w_r_1}) gives
\begin{equation}
\begin{aligned}
\rho_S\approx\frac{1}{N}+\frac{k\theta}{N}\left( \frac{R s\left(N-2k\right) }{2k\left(1 - \sum_{\alpha=1}^{n} p_{\alpha}^2\right)}-\frac{c\left(N-2\right)}{2}\right)
\end{aligned}.
\label{fix_u_w_r_2}
\end{equation}
Moreover, according to Eqs.~(\ref{exp_GE_fix}), (\ref{exp_GE_fix_1}) and (\ref{fix_u_w_r_2}), we have
\begin{equation}
\begin{aligned}
\mathbb{E}\left[E_{\mathbf{G}}\right]&= \frac{1}{2}\left(1+k\theta\left( \frac{R s\left(N-2k\right) }{2k\left(1 - \sum_{\alpha=1}^{n} p_{\alpha}^2\right)}-\frac{c\left(N-2\right)}{2}\right)\right)\left(\sum_{i=1}^{N}\frac{1-\sum_{\alpha=1}^{n}p_{\alpha}^{2}}{N s\left(k+1\right)}-\sum_{i=1}^{N}\frac{1-\sum_{\alpha=1}^{n}p_{\alpha}^{2}}{N s}\right)+\sum_{i=1}^{N}\frac{1-\sum_{\alpha=1}^{n}p_{\alpha}^{2}}{N s}\\&=\frac{1}{2}\left(1+k\theta\left( \frac{R s\left(N-2k\right) }{2k\left(1 - \sum_{\alpha=1}^{n} p_{\alpha}^2\right)}-\frac{c\left(N-2\right)}{2}\right)\right)\left(\frac{1-\sum_{\alpha=1}^{n}p_{\alpha}^{2}}{s\left(k+1\right)}-\frac{1-\sum_{\alpha=1}^{n}p_{\alpha}^{2}}{s}\right)+\frac{1-\sum_{\alpha=1}^{n}p_{\alpha}^{2}}{s}\\&
=-\frac{k\left(1-\sum_{\alpha=1}^{n}p_{\alpha}^{2}\right)}{2s\left(k+1\right)}\left(1+k\theta\left( \frac{R s\left(N-2k\right) }{2k\left(1 - \sum_{\alpha=1}^{n} p_{\alpha}^2\right)}-\frac{c\left(N-2\right)}{2}\right)\right)+\frac{1-\sum_{\alpha=1}^{n}p_{\alpha}^{2}}{ s}\\&
=\frac{1-\sum_{\alpha=1}^{n}p_{\alpha}^{2}}{s}\left(1-\frac{k}{2\left(k+1\right)}\left(1+k\theta\left( \frac{R s\left(N-2k\right) }{2k\left(1 - \sum_{\alpha=1}^{n} p_{\alpha}^2\right)}-\frac{c\left(N-2\right)}{2}\right)\right)\right).
 \label{exp_GE_fix_regular_homo}
\end{aligned}
\end{equation}

\section{Optimal degree for maximizing the collective performance}
Through our above theoretical derivation, we emphasize that our analytical framework can be applied to identify the network that is most favorable for group performance. By computing $\mathbb{E}\left[E_{\mathbf{G}}\right]$ for each graph $\mathbf{G}$ in the graph set $\bm{\mathcal{G}}$, the graph with the minimum $\mathbb{E}\left[E_{\mathbf{G}}\right]$ can be determined, corresponding to the optimal network for promoting collective performance. 

Furthermore, we find that  the expected average individual error on regular graphs depends on the degree $k$ through Eq.~(\ref{exp_GE_fix_regular_homo}). This indicates that the degree $k$ plays a significant role in determining the collective performance in the networks. We further provide an analytical derivation to explore the optimal $k^{*}$  maximizing the collective performance. Specifically, based on Eq.~(\ref{exp_GE_fix_regular_homo}), we define 
\begin{equation}
    h\left(x\right)=\frac{1-\sum_{\alpha=1}^{n}p_{\alpha}^{2}}{s}\left(1-\frac{x}{2\left(x+1\right)}\left(1+x\theta\left( \frac{R s\left(N-2x\right) }{2x\left(1 - \sum_{\alpha=1}^{n} p_{\alpha}^2\right)}-\frac{c\left(N-2\right)}{2}\right)\right)\right).
\end{equation}
The derivative of $h\left(x\right)$ with respect to $x$ is
\begin{equation}
    \frac{\partial h}{\partial x}=\frac{\left(2 R s+\left(1-\sum_{\alpha=1}^{n}p_{\alpha}^{2}\right)\left(N-2\right)c\right)\theta x^2+ 2\left(2 R s+ \left(1-\sum_{\alpha=1}^{n}p_{\alpha}^{2}\right) c \left(N-2\right) \right)\theta x-2 \left(1-\sum_{\alpha=1}^{n}p_{\alpha}^{2}\right)-N R s \theta}{4 s \left(x+1\right)^2}.
\end{equation}
Solving $\frac{\partial h}{\partial x}=0$ yields two roots, $x_1$ and $x_2$, which are respectively given by  
\begin{equation}
    x_1= - \sqrt{\frac{ (N + 2) R s \theta + 
    \left(1-\sum_{\alpha=1}^{n}p_{\alpha}^{2}\right) ((N-2)c \theta +2)}{\left(2 R s+\left(1-\sum_{\alpha=1}^{n}p_{\alpha}^{2}\right)\left(N-2\right)c\right)\theta}}-1=- \sqrt{\frac{ N R s \theta + 
    2\left(1-\sum_{\alpha=1}^{n}p_{\alpha}^{2}\right) }{\left(2 R s+\left(1-\sum_{\alpha=1}^{n}p_{\alpha}^{2}\right)\left(N-2\right)c\right)\theta}+1}-1,
\end{equation}
and 
\begin{equation}
    x_2= \sqrt{\frac{ (N + 2) R s \theta + 
    \left(1-\sum_{\alpha=1}^{n}p_{\alpha}^{2}\right) ((N-2)c \theta +2)}{\left(2 R s+\left(1-\sum_{\alpha=1}^{n}p_{\alpha}^{2}\right)\left(N-2\right)c\right)\theta}}-1=\sqrt{\frac{ N R s \theta + 
    2\left(1-\sum_{\alpha=1}^{n}p_{\alpha}^{2}\right) }{\left(2 R s+\left(1-\sum_{\alpha=1}^{n}p_{\alpha}^{2}\right)\left(N-2\right)c\right)\theta}+1}-1.
\end{equation}
Moreover, we have 
\begin{equation}
   \begin{cases}\frac{\partial h}{\partial x}>0 & x\in\left(-\infty,x_1\right)\cup\left(x_2,+\infty\right),\\
   \frac{\partial h}{\partial x}<0 & x\in\left(x_1,x_2\right).
   \end{cases}
\end{equation}
Since $x_1<0$ and $x_2>0$ always hold, we have 
\begin{equation}
\min_{x>0} h\left(x\right)=h\left(x_2\right)=h\left(\sqrt{\frac{ N R s \theta + 
    2\left(1-\sum_{\alpha=1}^{n}p_{\alpha}^{2}\right) }{\left(2 R s+\left(1-\sum_{\alpha=1}^{n}p_{\alpha}^{2}\right)\left(N-2\right)c\right)\theta}+1}-1\right).    
\end{equation}
Given that the degree must be an integer, we analytically determine the optimal $k^{*}$, which is 
\begin{equation}
k^{*} \in \left\{ \left\lfloor \sqrt{\frac{ N R s \theta + 
    2\left(1-\sum_{\alpha=1}^{n}p_{\alpha}^{2}\right) }{\left(2 R s+\left(1-\sum_{\alpha=1}^{n}p_{\alpha}^{2}\right)\left(N-2\right)c\right)\theta}+1}-1 \right\rfloor, \left\lceil \sqrt{\frac{ N R s \theta + 
    2\left(1-\sum_{\alpha=1}^{n}p_{\alpha}^{2}\right) }{\left(2 R s+\left(1-\sum_{\alpha=1}^{n}p_{\alpha}^{2}\right)\left(N-2\right)c\right)\theta}+1}-1 \right\rceil \right\},
\end{equation}
where $\lfloor \cdot \rfloor$ and $\lceil \cdot \rceil$ denote the floor and ceiling functions, respectively. The value of $k^*$ is chosen as the one that minimizes the expected average individual error while satisfying the structural requirement of a regular graph, i.e., $k^{*}\geq 2$.

Moreover, for sufficiently large $N$, we have
\begin{equation}
    x_2 \rightarrow \sqrt{\frac{ R s}{\left(1-\sum_{\alpha=1}^{n}p_{\alpha}^{2}\right)c}+1}-1.
\end{equation}
Accordingly, the optimal $k^{*}$ is approximated as
\begin{equation}
k^{*} \in \left\{ \left\lfloor \sqrt{\frac{ R s}{\left(1-\sum_{\alpha=1}^{n}p_{\alpha}^{2}\right)c}+1}-1 \right\rfloor, \left\lceil \sqrt{\frac{ R s}{\left(1-\sum_{\alpha=1}^{n}p_{\alpha}^{2}\right)c}+1}-1 \right\rceil \right\}.
\end{equation}

\clearpage
\bibliography{reference}
\clearpage
\subsection*{Supplementary Figures}

\begin{figure*}[!h]
\centering
\includegraphics[width=16cm]{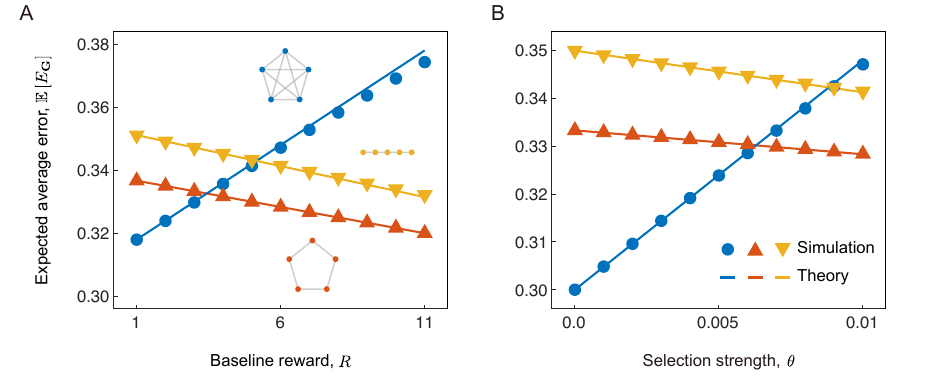}
\caption{\textbf{Expected average error on three typical five-node graphs.} As shown in the main text, complete, ring, and line graphs are representative five-node graphs for collective performance. We show the average individual error as a function of baseline reward $R$ (A) and selection strength $\theta$ (B) across these graphs, highlighting how the two factors influence collective performance on networks. In the complete graph, group performance is high when the baseline reward (A) or selection strength (B) is low; increasing either parameter raises the average individual error, thereby reducing collective performance. In ring and line graphs, however, the average individual error is higher than that in the complete graph when the baseline reward (A) or selection strength (B) is small. As either parameter increases, the average individual error on these two graphs decreases, leading to an improvement in group performance. Our theoretical predictions (lines) are in agreement with numerical simulations (dots). We set $\theta=0.01$ in (A) and $R=6$ in (B), other parameters are the same as those in Fig.~1A.}
\label{SI_n_5_R_Delta}
\end{figure*}

\clearpage 

\begin{figure*}
\centering
\includegraphics[width=16cm]{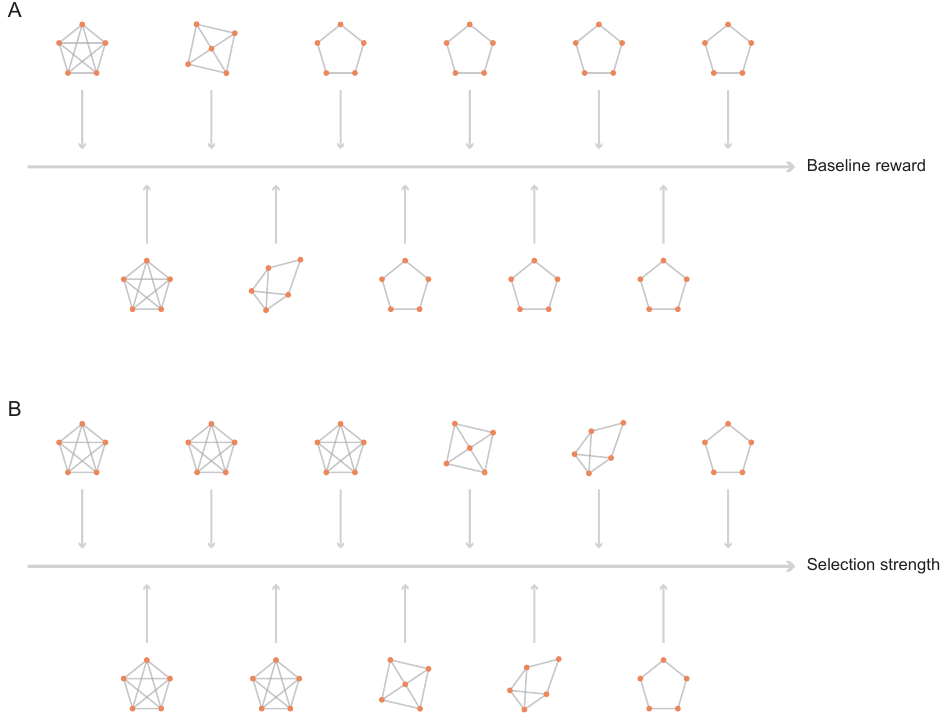}
\caption{\textbf{Effect of selection strength and baseline reward on optimal five-node graphs for collective performance.} We show the optimal graphs of size five with increasing baseline reward (A) and selection strength (B), revealing a transition of the optimal graph from a complete graph to a ring graph. In (A), we set $\theta=0.01$ and increase baseline reward $R$ from $1$ to $11$ with interval $1$. In (B), we set $R=6$ and increase selection strength $\theta$ from $0$ to $0.01$ with interval $0.001$. Other parameters are the same as those in Fig.~1A.}
\label{SI_n_5_optimal}
\end{figure*}

\clearpage 

\begin{figure*}
\centering
\includegraphics[width=16cm]{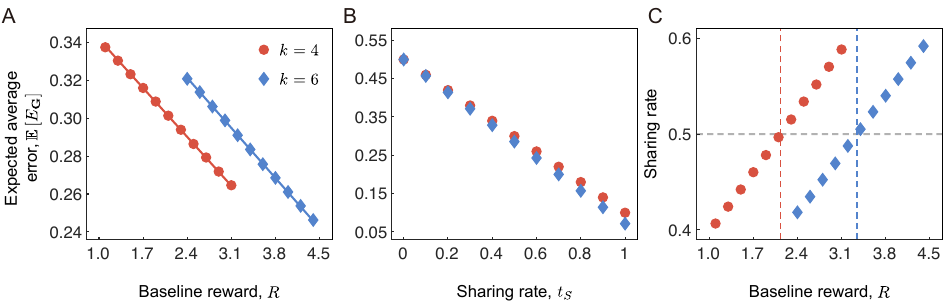}
\caption{\textbf{Collective performance on random regular graphs} (A) Average individual error as a function of baseline reward $R$ in random regular graphs with different average degrees ($k=4$, red; $k=6$, blue). Our theoretical predictions (lines) are in agreement with numerical simulations (symbols). Although networks with higher degrees generally facilitate information transmission, those with a lower average degree ($k = 4$) achieve superior collective performance when strategy evolution is considered. Each data point is obtained through $10^8$ Monte Carlo simulations. (B) Average individual error across networks with different average degrees as a function of sharing rate. At a given sharing rate, networks with higher average degree consistently achieve better collective performance. Moreover, higher sharing rate further enhances collective performance. (C) Information sharing rate as a function of baseline reward $R$. For a given baseline reward, networks with a lower average degree exhibit higher sharing rates, suggesting that sparser networks more effectively promote the evolutionary advantage of the information sharing strategy. Theoretical predictions also identify the critical baseline reward at which sharing is favored. Parameters: $N=100$, $\theta=0.002$, $n=2$, $p_1=p_2=\frac{1}{2}$, and $s_m=c_m=1$ for $m=1,\cdots,N$.}
\label{SI_RR}
\end{figure*}

\clearpage 

\begin{figure*}
\centering
\includegraphics[width=16cm]{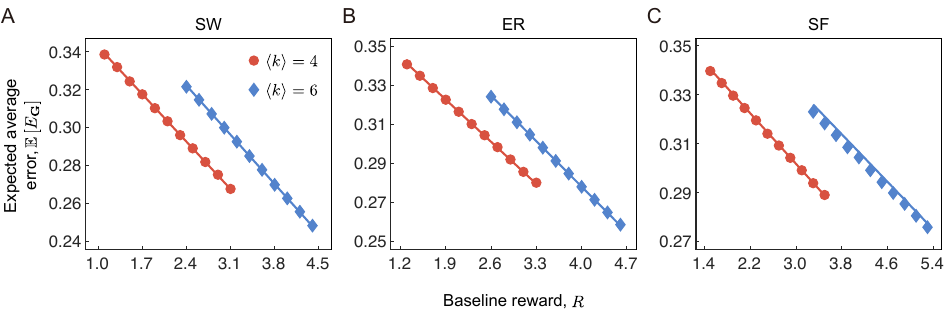}
\caption{\textbf{Monte Carlo simulations of collective performance across different networks.} (A-C) Average individual error as a function of baseline reward across three classes of networks: Watts--Strogatz (SW) small–world networks with rewiring probability 0.1, Erd\H{o}s--R\'enyi (ER) random networks, and Barab\'asi--Albert scale-free (SF) networks (from left to right). Theoretical predictions (lines) are in agreement with numerical simulations (symbols), demonstrating the robustness of our analytical framework across diverse network topologies. Other settings and parameters are consistent with those in Fig.~S3.}
\label{SI_other_structure}
\end{figure*}

\clearpage 

\begin{figure*}
\centering
\includegraphics[width=16cm]{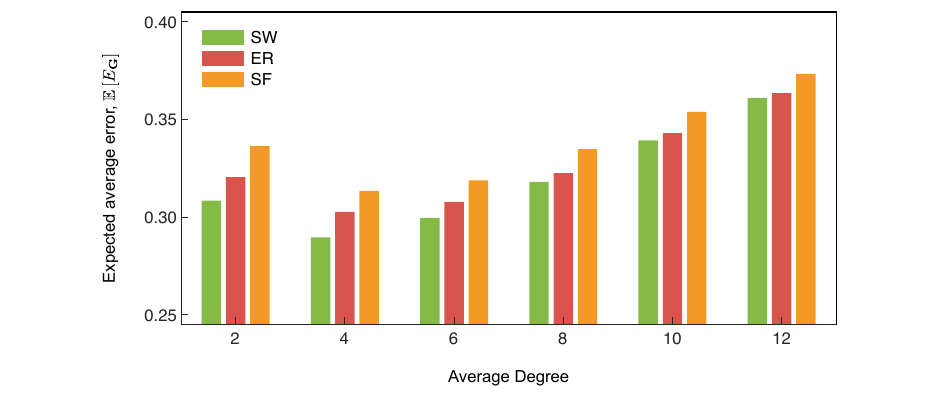}
\caption{\textbf{Effect of average degree on collective performance across different networks.} We show the average individual error with increasing average degree across Watts--Strogatz small–world (SW) networks with rewiring probability 0.1, Erd\H{o}s--R\'enyi  random (ER) networks, and Barab\'asi--Albert scale-free (SF) networks. A consistent non-linear effect of average degree on collective performance is observed across these networks, where an intermediate degree maximizes group performance by minimizing the average individual error. For each network type and average degree, the expected average error is obtained as the mean over $20$ independently generated networks. Other settings and parameters are consistent with those in Fig.~3.}
\label{SI_optimal_other_structure}
\end{figure*}

\clearpage 

\begin{figure*}
\centering
\includegraphics[width=16cm]{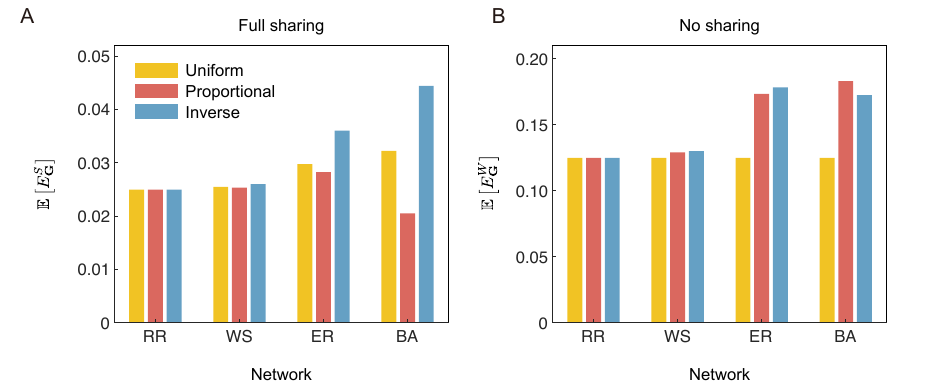}
\caption{\textbf{Effect of sample allocation on collective performance in homogeneous states.} In the absence of strategy evolution, we simulate the average individual error under the full-sharing (A) and non-sharing (B) states across four network types: random regular (RR) graphs, Watts–Strogatz small-world (SW) networks with a rewiring probability of $0.1$, Erdős–Rényi random (ER) networks, and Barabási–Albert scale-free (SF)  networks. The results show that, in both homogeneous states without strategy evolution, allocating samples inversely proportional to node degree does not maximize collective performance among the three allocation patterns (i.e., uniform, proportional, and inverse). All other settings and parameters are consistent with those used in Fig.~4.}
\label{SI_allocation}
\end{figure*}

\clearpage 

\begin{figure*}
\centering
\includegraphics[width=16cm]{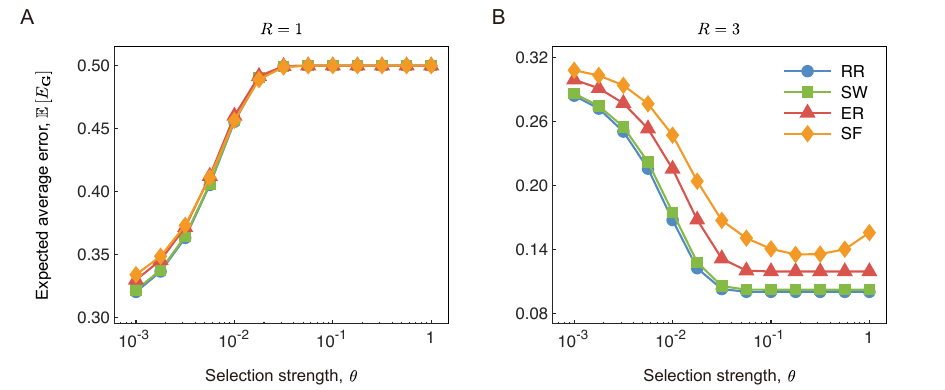}
\caption{\textbf{Double-sided role of selection strength in collective performance.} We simulate the average individual error with increasing selection strength across random regular (RR) graphs, Watts--Strogatz small–world (SW)  networks with rewiring probability 0.1, Erd\H{o}s--R\'enyi random (ER) 
 networks, and Barab\'asi--Albert scale-free (SF) networks. We show that the effect of selection strength on collective performance depends on the baseline reward $R$. (A) For a low baseline reward ($R=1$), increasing selection strength raises the expected average individual error, thereby reducing collective performance. (B) For a higher baseline reward ($R=3$), increasing selection strength lowers the average individual error, thus enhancing collective performance. In more heterogeneous networks, such as the scale-free (SF) network, the average individual error initially decreases with increasing selection strength but rises as $\theta \rightarrow 1$. However, this error under strong selection remains lower than that under neutral drift ($\theta=0$), in which strategy evolution does not influence group performance. Other settings and parameters are consistent with those in Figs.~S3-4.}
\label{SI_strong_selection}
\end{figure*}

